\documentclass[structabstract]{aa} 
\usepackage{graphicx}
\usepackage{subfig}
\usepackage{longtable}
\usepackage{footmisc}
\usepackage{amsmath}
\usepackage{txfonts}
\usepackage{upgreek}
\usepackage{natbib}
\bibpunct{(}{)}{;}{a}{}{,}

\newcommand{\revone}[1]{\textnormal{#1}}
\newcommand{\revtwo}[1]{\textnormal{#1}}

\begin{document}
   \title{VLT imaging of the $\beta\,$Pictoris gas disk\thanks{Based on observations made with ESO Telescopes at the La Silla Paranal Observatory under programme IDs 382.C-0394 and 384.C-0551}}


   \author{R.\ Nilsson
          \inst{1}
          \and
           A.\ Brandeker
          \inst{1}
          \and       
           G.\ Olofsson
          \inst{1}
	 	  \and
	  	  K.\ Fathi
		  \inst{1}
		  \and
          Ph.\ Th\'ebault
          \inst{2}
          \and
           R.\ Liseau
          \inst{3}
          }
          
   \offprints{R.\ Nilsson}

   \institute{Department of Astronomy, Stockholm University, AlbaNova University Center, Roslagstullsbacken 21, SE-106 91 Stockholm, Sweden\\
              \email{ricky@astro.su.se, alexis@astro.su.se, olofsson@astro.su.se, kambiz@astro.su.se}
	\and
	   Observatoire de Paris, Section de Meudon, F-92195 Meudon Principal Cedex, France\\
	   \email{philippe.thebault@obspm.fr}
         \and
             Onsala Space Observatory, Chalmers University of Technology, SE-439 92 Onsala, Sweden\\
             \email{rene.liseau@chalmers.se}
         		 }

   \date{Received March 27, 2012; accepted July 3, 2012}

 
  \abstract
   {Circumstellar debris disks older than a few Myr should be largely devoid of primordial gas remaining from the protoplanetary disk phase. Tracing the origin of observed atomic gas in Keplerian rotation in the edge-on debris disk surrounding the $\sim$12\,Myr old star $\beta\,$Pictoris requires more detailed information about its spatial distribution than has previously been acquired by limited slit spectroscopy. Especially indications of asymmetries and presence of \ion{Ca}{ii} gas at high disk latitudes call for additional investigation to exclude or confirm its connection to observed dust structures or suggested cometary bodies on inclined eccentric orbits.}
   {We set out to recover a complete image of the \ion{Fe}{i} and \ion{Ca}{ii} gas emission around $\beta\,$Pic by spatially resolved, high-resolution spectroscopic observations to better understand the morphology and origin of the gaseous disk component.}
   {The multiple fiber facility FLAMES/GIRAFFE at the Very Large Telescope (VLT), with the large integral-field-unit ARGUS, was used to obtain spatially resolved optical spectra (from 385.9 to 404.8\,nm) in four regions covering the northeast and southwest side of the disk. Emission lines from \ion{Fe}{i} (at 386.0\,nm) and \ion{Ca}{ii} (at 393.4 and 396.8\,nm) were mapped and could be used to fit a parametric function for the disk gas distribution, using a gas-ionisation code for gas-poor debris disks.}
   {Both \ion{Fe}{i} and \ion{Ca}{ii} emission are clearly detected, with the former dominating along the disk midplane, and the latter revealing vertically more extended gas. The surface intensity of the \ion{Fe}{i} emission is lower but more extended in the northeast (reaching the 210\,AU limit of our observations) than in the southwest, while \ion{Ca}{ii} shows the opposite asymmetry. The modelled Fe gas disk profile shows a linear increase in scale height with radius, and a vertical profile that suggests dynamical interaction with the dust. We also qualitatively demonstrate that the \ion{Ca}{ii} emission profile can be explained by optical thickness in the disk midplane, and does not require Ca to be spatially separated from Fe.}
   {}

	 \titlerunning{VLT imaging of the $\beta\,$Pictoris gas disk}

   \maketitle
%

\section{Introduction}

The transformation of dusty and gaseous disks around young stars into planetary systems involves a gradual clearing of the disk by several processes. Dense and gas-rich protoplanetary envelopes are affected by strong stellar radiation \citep{Gorti2009} and winds \citep{Lovelace2008}, that compete with viscous accretion \citep{Lynden-Bell1974} to open up growing inner disk holes, and cause complete photoevaporation of the gaseous disk component within $\sim$10\,Myr \citep{Zuckerman1995,Haisch2001,Jayawardhana2006,Pascucci2006}. This also sets the time frame for giant planet formation, because it restricts the amount of gas available to be swept up by growing planetary embryos. Primordial dust grains are blown out from the system or spiral onto the star due to Poynting-Robertson (PR) drag and are cleared from the disk within a few thousand years. The left-over debris disk is defined by a low ($<$10\%) gas-to-dust ratio, is mostly optically thin, and often observable as excess mid- and far-infrared (IR) emission in the spectral energy distribution (SED), originating from heated dust grains that are being continuously produced in collisions between planetesimals \citep{Dominik2003,Wyatt2007}. Decrease in this steady-state collisional processing and continous removal of submicron sized dust by radiation pressure leads to a decline in observed IR excess with time. Most of the dust in a standard debris disk is believed to be composed of porous silicate grains containing Si, Mg, and Fe, likely combined with oxygen into forsterite (Mg$_{2}$SiO$_{4}$), enstatite (MgSiO$_{3}$), olivine ([Mg,Fe]$_{2}$SiO$_{4}$), and pyroxenes ([Mg,Fe]SiO$_{3}$), as found in chondritic meteorites \citep[see review by][]{Henning2010}. Beyond the ice condensation (snow) line, grains would contain a cometary fraction of H$_{2}$O, CO$_{2}$, CO, NH$_{3}$, CH$_{4}$, and N$_{2}$ ices, although there is a possibility that photosputtering might strip grains from their ices beyond the snowline \citep{Grigorieva2007}.

Observation of gas in such evolved non-primordial disks is not expected a priori. It is of course routinely observed in younger systems in the protoplanetary-disk stage. As an example, abundant molecular species (e.g.\ CO and H$_{2}$) in young circumstellar disks have been detected in both emission \citep{Najita2003,Herczeg2006} and absorption \citep{Redfield2007,Roberge2008}. However, in most disk environments, especially around early-type stars, molecules like H$_{2}$O, OH, and CO are not expected to survive, instead producing an abundance of atomic oxygen and carbon by photodissociation \citep[e.g.][]{Kamp2000}. Elements with strong resonant transitions in the ultraviolet (UV) and optical, especially metals, are then easily blown out from the system by radiation pressure. Thus, the detection of substantial quantities of gas in older systems that have presumably made the transition to debris disks, like the 30\,Myr-old HD\,21997 \citep{Moor2011} and 200\,Myr-old $\sigma$~Her \citep{Chen2003}, is surprising and points to unknown gas retention or production mechanisms. If gas is being produced, it can give us information on the composition of solid bodies in the disk, and even small amounts of gas can affect the dynamics of dust grains \citep{Thebault2005, Krivov2009}. $\beta\,$Pictoris is the only currently known debris disk that contains gas and is close enough to be spatially resolved at optical and IR wavelengths, allowing detection of resonantly scattered light and thus a more detailed study of gas composition and distribution.


The first indication of circumstellar material around the $\sim$12\,Myr A6V star $\beta\,$Pic came from detection of thermal IR excess by the \emph{Infrared Astronomical Satellite} \citep[IRAS,][]{Aumann1985}, and subsequent ground-based coronagraphic imaging revealed a flared, nearly edge-on dust disk in scattered optical and near-IR light \citep{Smith1984, Parasce1987}, extending out to 1800\,AU. Following its classification as a debris disk star \citep[or so-called Vega-like star, from the prototype IR excess star Vega,][]{Backman1993}, $\beta\,$Pic has been extensively studied and the \revone{minimum mass of cold dust} contained within its disk determined from submillimeter (submm) and millimeter observations \revone{to 3--10 lunar masses} \citep{Zuckerman1993,Holland1998,Dent2000,Liseau2003,Nilsson2009}. 
In addition to the collisionally produced dust \citep{Backman1993,Artymowicz1997,Lagrange2000,Zuckerman2001}, several lines of evidence suggest that evaporating comets on eccentric orbits, so-called \emph{falling evaporating bodies}, {FEBs}, are supplying dust to the $\beta\,$Pic debris disk \citep{LecavelierDesEtangs1996,LecavelierDesEtangs1998b}. One comes from spectroscopic signatures of intermittent redshifted absorption lines \citep[][and references therein]{Lagrange-Henri1988,Beust1990,Vidal-Madjar1994} and a second from emission by crystalline silicates (at 9.7, 28, and 33.5$\,\mu$m) and olivines (at 11.3$\,\mu$m) \citep{Telesco1991,Knacke1993,Aitken1993,Weinberger2003,Okamoto2004,Chen2007}, similar to that observed in e.g.\ comet Halley. This points to a cometary-like grain composition of either crystalline olivine ($\sim$55\%), pyroxene ($\sim$35\%), and other silicates, or a mix of 95\% amorphous and 5\% crystalline olivine \citep[][and references therein]{Artymowicz1997}. Dissimilarities between disk features in IR emission and scattered light images of the disk imply that two distinctly different grain populations contribute, perhaps low-albedo refractory grains (like the hot, $>300\,$K, silicates that are unmistakably responsible for the 10$\,\mu$m emission) and high-albedo icy grains.

Information about the spatial distribution of circumstellar gas is more limited than the dust, since it is more difficult to observe. Several species of atomic gas \revone{have} been detected in absorption, due to the disk's favorable edge-on orientation. The strong circumstellar \ion{Ca}{ii}\,K absorption line, e.g., was detected early, together with weaker \ion{Na}{i}\,D lines \citep{Vidal-Madjar1986,Hobbs1985}, and several neutral and singly ionised metallic species have been observed in both optical and UV absorption spectroscopy \citep{Lagrange1998}. Surprisingly, these appear to be very stable (apart from the intermittent redshifted components) and do not show any radial velocity, which would be expected from acceleration by radiation pressure. 

\citet{Olofsson2001} performed high-resolution slit spectroscopy along the disk of $\beta\,$Pic and made the first spatially resolved detection of circumstellar atomic gas around any star. The \ion{Na}{i}\,D emission line (doublet) at 5895\,{\AA} was found to be resonantly scattered from 30 to 140\,AU, with an average \ion{Na}{i} column density of $10^{15}\,$cm$^{-2}$ and a velocity pattern clearly indicating gas in Keplerian rotation. To explain the absence of any detected radial motion -- considering the fact that the force from radiation pressure felt by \ion{Na}{i} throughout the $\beta\,$Pic disk is 250 times stronger than the gravitational force -- \citet{Olofsson2001} proposed fast ionisation by the intense stellar UV radiation, which does not allow significant time for acceleration of neutral \ion{Na}{i} before conversion to \ion{Na}{ii} ions (which lack strong transitions in the wavelength region dominating the stellar SED). This model was recently improved by \citet{Brandeker2011}. In \emph{VLT/UVES} observations over a wider spectral range and with a spatial and spectral resolution twice as high, \citet{Brandeker2004} detected 88 emission lines originating from \ion{Fe}{i}, \ion{Na}{i}, \ion{Ca}{ii}, \ion{Ni}{i}, \ion{Ni}{ii}, \ion{Ti}{i}, \ion{Ti}{ii}, \ion{Cr}{i}, and \ion{Cr}{ii}. Spatial information from the slit placed at four positions across the disk (at 60 and 120\,AU, on either side) and four positions along the disk revealed a surprisingly extended and asymmetric gas distribution. The strong \ion{Fe}{i} and \ion{Na}{i} emission reaches along the disk from 13 out to 323\,AU from the star, displaying a NE-SW asymmetry similar to that observed in the dust, with a gradual NE radial decline and a sharp cutoff in the brighter (out to 75\,AU) SW emission at 150--200\,AU. Also in accordance with dust observations, an inner 5\degr\ tilt was found by \citet{Brandeker2004}. The radial thickness increase of the gas disk appears to be significantly higher than that of the dust (almost twice the scale-height of the dust disk at 116\,AU). \ion{Ca}{ii} H- and K emission can be traced to heights of 77\,AU above the mid-plane, but falls off closer to the mid-plane. As in the observations by \citet{Olofsson2001}, the outward radial motion of gas is much slower than would be expected from acceleration due to radiation pressure. This points to a braking agent \revtwo{that cannot be \ion{H}{i} and H$_{2}$, which} have observational upper limits in column density inconsistent with observed velocities, or perhaps an unknown gas production mechanism \citep{Brandeker2004}. \citet{Fernandez2006} studied gas braking and production processes, concluding that deceleration of ions through Coulomb interaction with other ions (self-braking) could be sufficient to brake the gas, but only if C in the disk gas is at least five times overabundant compared to solar elemental abundances. Spectroscopic observations with the \emph{Far Ultraviolet Spectroscopic Explorer} (FUSE) \citep{Roberge2006} did find \ion{C}{ii} and \ion{C}{iii} measured in absorption to be significantly more abundant compared to elemental abundances of solar system material (viz.\ the Sun, carbonaceous chondrites, and dust from comet Halley), and could thus partly solve the radial velocity conundrum. More detailed modelling of the braking by \citet{Brandeker2011} and observations by the PACS instrument at the \textit{Herschel Space Observatory} (Brandeker et al.\ 2012) indicated that the C abundance over metallic elements such as Na and Fe might be even higher, of about 400$\times$ solar abundance.

Because of gas-removal mechanisms, the gas is probably not primordial \citep{Fernandez2006}, but replenished. Suggested mechanisms include (1) gas released from star-grazing comets \citep{Beust2007}, (2) photo-desorption from circumstellar dust grains \citep{Chen2007}, and (3) vaporisation of dust by collisions with high-velocity $\beta$-meteoroids \citep{Czechowski2007}. Since these different mechanisms should produce different spatial distributions of gas, one way to potentially distinguish between them is to map the spatial distribution of gas, which is the aim of this study. One would, e.g., expect most of the gas released by comets to come from the inner regions of the system; the photo-desorption release of gas would be expected to be $\propto A(r)/r^2$, where $r$ is the distance to the star and $A(r)$ is the dust area density (since the UV radiation responsible for photo-desorption drops as $r^2$); while the dust-dust collision gas production profile would need more detailed modelling \citep[as shown by][]{Czechowski2007}. Once produced, the gas may get distributed/removed, which is another effect that needs to be taken into account to properly predict the current spatial distribution of gas. This detailed modelling is beyond the scope of the present paper, however, which focuses on the observed gas profile. 

To obtain a complete map of the \ion{Fe}{i} (at 386.0\,nm) and \ion{Ca}{ii} (at 393.4 and 396.8\,nm) emission throughout the disk of $\beta\,$Pic, we used high-resolution integral field spectroscopy to cover the disk (Sects.~\ref{sec:obs} and \ref{sec:res}). The observed emission from \ion{Fe}{i} is modelled to obtain the spatial number density distribution of gas, combining a de-projection technique (Sect.~\ref{subsec:fit}) with the ionisation and thermal equilibrium code \textsc{ontario} (Sect.\ref{subsec:ont}). The obtained gas distribution is then used to compute the expected shape of the \ion{Ca}{ii} H emission profile, using a simple radiative transfer code to handle the optically thick H line (Sect.~\ref{subsec:caiimodel}). Our conclusions are summarised in Sect.~\ref{sec:con}.



\section{Observations and data reduction}\label{sec:obs}

\subsection{Observations with FLAMES/GIRAFFE at VLT}

The multi-object fiber facility \emph{FLAMES} at the 8.2-m diameter Kueyen (UT2) telescope of the \emph{Very Large Telescope} (VLT) array was used to obtain spatially resolved, high spectral resolution optical data of the $\beta\,$Pic disk, employing its \emph{GIRAFFE} spectrograph with the \emph{ARGUS} fiber configuration. ARGUS is a large integral-field-unit (IFU) composed of a rectangular array of $22{\times}14$ microlenses, sampling either $0.52\arcsec$/microlens (with a total sky coverage of $11.5\arcsec{\times}7.3\arcsec$) or $0.30\arcsec$/microlens ($6.6\arcsec{\times}4.2\arcsec$), and feeding the spectrograph through individual fibers. The high-resolution (HR) dispersion grating (with $R$\,$\approx$\,31\,300 or $d\lambda$\,$\approx$\,$0.0126\,$nm) offers a total spectral range of 370--950\,nm accessed through various setups. We used the bluest HR setting with the central wavelength of the GIRAFFE grating at 395.8\,nm.
\begin{figure} 
\centering
   \includegraphics[trim=5mm 5mm 5mm 0mm, clip=true, width=0.48\textwidth]{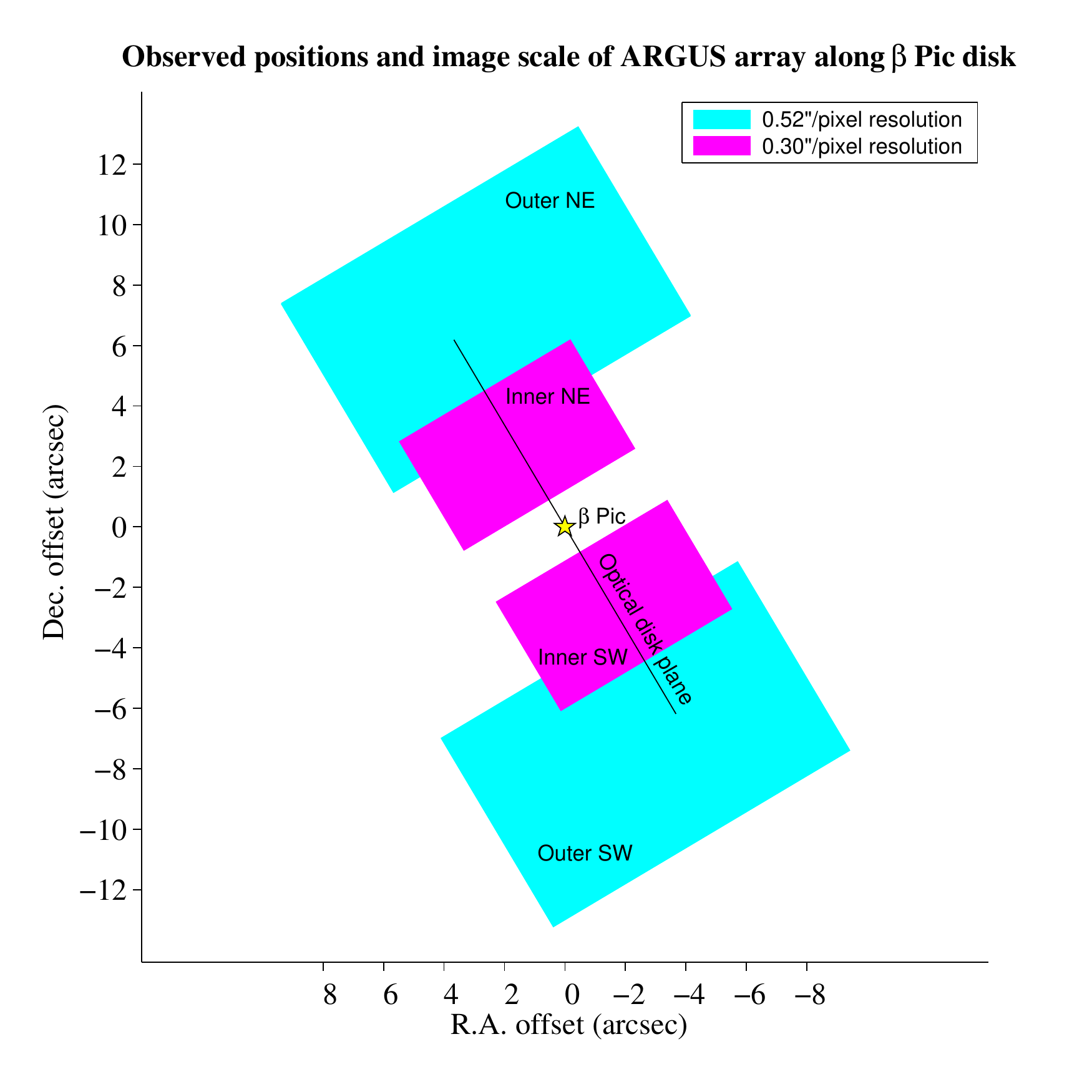}
      \caption{Scale and position in the sky of the ARGUS array during the observations. The northeast (NE) and southwest (SW) side of the optical disk plane (marked with a black line) was covered, omitting the area closest to the star, where the stellar PSF and scattered light would be strong, giving low signal-to-noise ratio. Cyan- and magenta-coloured areas represent the higher ($0.30\arcsec$/pixel) and lower ($0.52\arcsec$/pixel) spatial resolution settings used for the inner and outer disk, respectively.}
      \label{fig:arraypos}
\end{figure}
Observations were carried out in service mode, spread out over a period ranging from November 2008 to January 2010 (due to stringent $\lesssim0.5\arcsec$ seeing requirements), with an average airmass of 1.17. The ARGUS array was placed at four positions orthogonally along the disk (two on each side of the star), partially overlapping the smaller scale image of the inner disk with the larger scale image of the outer disk (as displayed in Fig.\ \ref{fig:arraypos}) to check for consistency. At each position, exposures were taken at two position angles (rotated $180\degr$) to reduce the risk of having any sky areas repeatedly falling on dead pixels. Table~\ref{table:obs} presents observing dates, field centre positions, position angles (PA), pixel scales, and exposure times of the observations.

\begin{table}
\caption{Observing log from VLT/FLAMES/GIRAFFE observations of $\beta\,$Pic. The optical coordinates of the star are $\textnormal{R.A.}=05\textnormal{h}\,47\textnormal{m}\,17.088\textnormal{s}$ and $\textnormal{Dec.}=-51\degr 03\arcmin 59.44 \arcsec$ (J2000).}             
\label{table:obs}      
\centering          
\begin{tabular}{ l l l l l l }     
\hline\hline       
RJD & R.A.\tablefootmark{a} & Dec.\tablefootmark{a} & PA & Scale & Exp.\ time\\ 
\hline
 & ($\degr$) & ($\degr$) & ($\degr$) & ($\arcsec$/pix) & (s) \\ 
\hline                    
   54774.72 & 86.8219 & -51.0656 & 300.75 & 0.30 & 895.0 \\  
   54774.73 & 86.8219 & -51.0656 & 300.75 & 0.30 & 895.0 \\ 
   54774.75 & 86.8219 & -51.0656 & 300.75 & 0.30 & 895.0 \\ 
   54774.82 & 86.8219 & -51.0656 & 120.75 & 0.30 & 895.0 \\ 
   54774.83 & 86.8219 & -51.0656 & 120.75 & 0.30 & 895.0 \\
   54774.84 & 86.8219 & -51.0656 & 120.75 & 0.30 & 895.0 \\
   54798.70 & 86.8205 & -51.0670 & 300.75 & 0.30 & 895.0 \\
   54798.71 & 86.8205 & -51.0670 & 300.75 & 0.30 & 895.0 \\
   54798.72 & 86.8205 & -51.0670 & 300.75 & 0.30 & 895.0 \\
   54799.69 & 86.8205 & -51.0670 & 120.75 & 0.30 & 895.0 \\
   54799.70 & 86.8205 & -51.0670 & 120.75 & 0.30 & 895.0 \\
   54799.71 & 86.8205 & -51.0670 & 120.75 & 0.30 & 895.0 \\
   54912.50 & 86.8224 & -51.0643 & 300.75 & 0.52 & 2775.0 \\
   54913.51 & 86.8224 & -51.0643 & 120.75 & 0.52 & 2775.0 \\ 
   55193.70 & 86.8200 & -51.0683 & 120.75 & 0.52 & 2775.0 \\
   55225.65 & 86.8200 & -51.0683 & 300.75 & 0.52 & 2775.0 \\
   55225.70 & 86.8224 & -51.0643 & 120.75 & 0.52 & 2775.0 \\
\hline                  
\end{tabular}
\tablefoot{
\tablefoottext{a}{ARGUS field centre position.}
}
\end{table}

\subsection{Reducing FLAMES/GIRAFFE data}\label{sec:reduction}

The raw images containing dispersed spectra from all fibers (including sky- and simultaneous calibration exposures) were reduced using ESO's recipe execution tool \emph{EsoRex} with supplied GIRAFFE pipeline recipes. Calibration products (master bias, master flat-field with fiber localisation data, dispersion solution and line identification from simultaneous ThAr arc lamp spectra, and instrument response from standard star observations) were created and then applied in the reduction of science frames to retrieve flux- and wavelength-corrected spectra in each pixel of the reconstructed image. The final spectra were extracted by summing up \revone{over all pixels in the fiber width and were subsequently rebinned to a linear scale of 0.005\,nm spectral elements (from the raw $\sim$0.0046\,nm/pixel non-linear scale), ranging from 385.9\,nm to 404.8\,nm, with a measured average resolution (from ThAr lines) full-width at half-maximum (FWHM) of 0.0111\,nm}. Line-flux errors were estimated from the noise in the region surrounding each line, but it should be noted that the absolute photometric accuracy of the instrument response at these blue wavelengths are not better than $\sim$20\%, judging from the standard deviation of obtained instrument response curves (see \ref{fig:instrumentresponse} in Appendix~\ref{sec:calibcomp}). 

The data cubes were further processed in \textsc{matlab}, where barycentric wavelength correction and airmass correction was applied, and images were combined. Most cosmic ray hits could be identified and eliminated by sigma clipping and vertical (spectral dimension) interpolation directly in the unprocessed images. Remaining spikes in the spectra of the inner disk could be removed by modified median filtering, replacing values in the contaminated spectrum with the mean of the two additional exposures obtained (after normalisation), but the long single exposures of the outer disk presented a bigger problem. In that region we had to replace identified spikes with values interpolated from neighbouring bins, taking care to correct for occasional adjacent or extraordinarily wide spikes by comparing with the rotated exposure. The fact that the unresolved emission lines we were mapping had a width comparable to the average width of the cosmic ray spikes added to the complication. 
%
%
%
\begin{figure} 
\centering
   \includegraphics[trim=35mm 4mm 0mm 0mm, clip=true, width=0.55\textwidth]{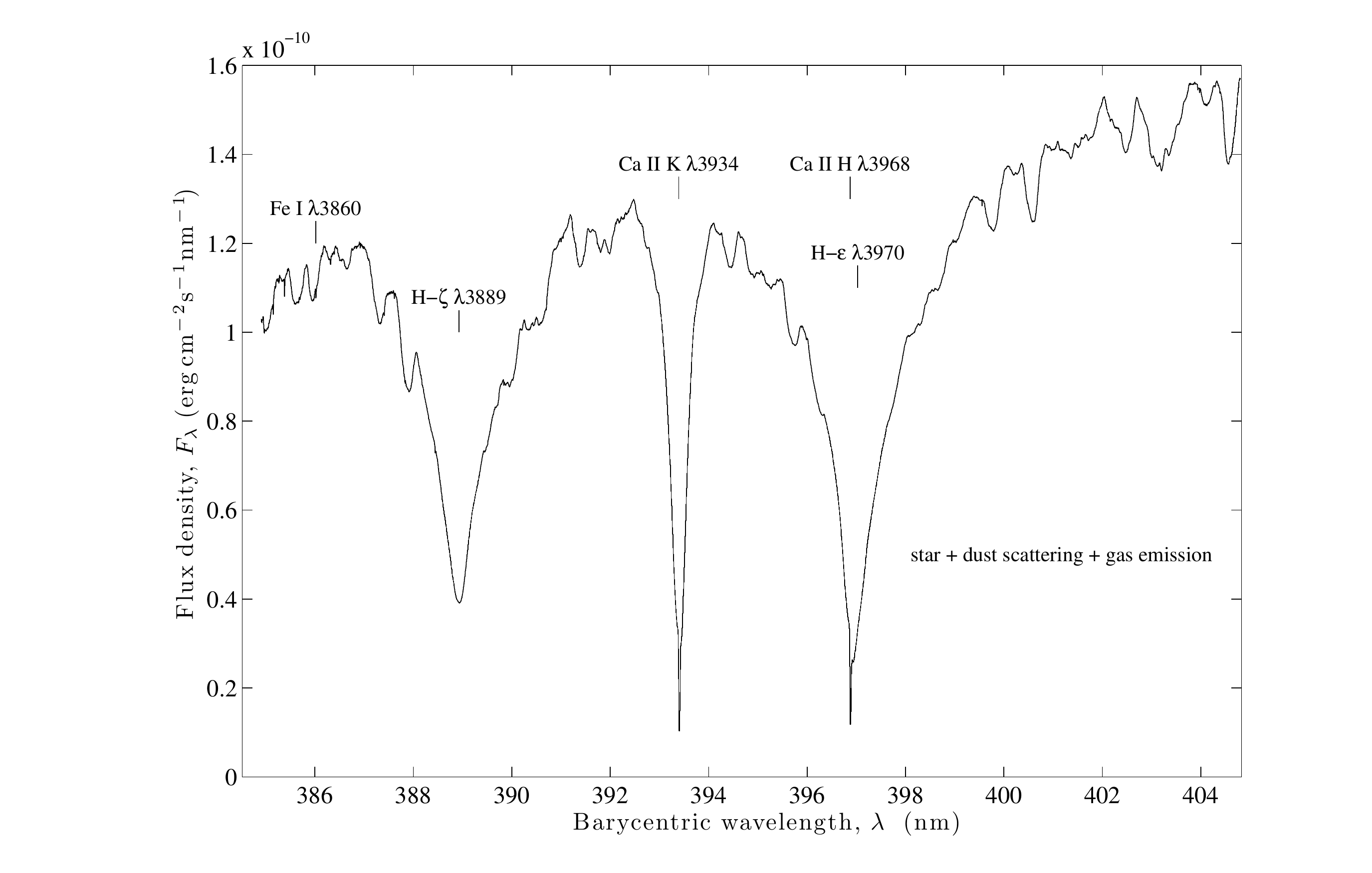}
      \caption{Total observed near-UV spectra of inner disk, comprised mainly of a stellar PSF component, with additions from gas emission and light scattered off circumstellar dust grains.}
      \label{fig:accfullspect}
\end{figure}
 
\subsection{Isolating the emission lines}

To extract the flux from the \ion{Fe}{i} and \ion{Ca}{ii} emission lines (Table~\ref{table:lines}) we had to subtract the dominant stellar point-spread function (PSF) and dust-scattered light contribution to the total spectrum (Fig.~\ref{fig:accfullspect}).

\begin{table}
\caption{Observed emission lines.}             
\label{table:lines}      
\centering          
\begin{tabular}{ l c c c c}     
\hline\hline       
Line & ${\lambda}_\textnormal{air}$ & $E_j{\rightarrow}E_i$ & $A_{ji}$ & $g_j-g_i$ \\ 
\hline
 & (nm) & (eV) & s$^{-1}$\\ 
\hline                    
   \ion{Fe}{i} & 385.99114 & $3.211{\rightarrow}0.000$ & $9.69 \times 10^{6}$ & $9-9$\\  
   \ion{Ca}{ii} K & 393.36614 & $3.151{\rightarrow}0.000$ & $1.47 \times 10^{8}$ & $2-4$ \\ 
   \ion{Ca}{ii} H & 396.84673 & $3.123{\rightarrow}0.000$ & $1.4  \times 10^{8}$ & $2-2$ \\ 
\hline                  
\end{tabular}
\tablefoot{Data retrieved from NIST (http://physics.nist.gov).}

\end{table}
\begin{figure*}  
\centering
	\subfloat[][]{\includegraphics[trim=35mm 4mm 17mm 0mm, clip=true, width=0.48\textwidth]{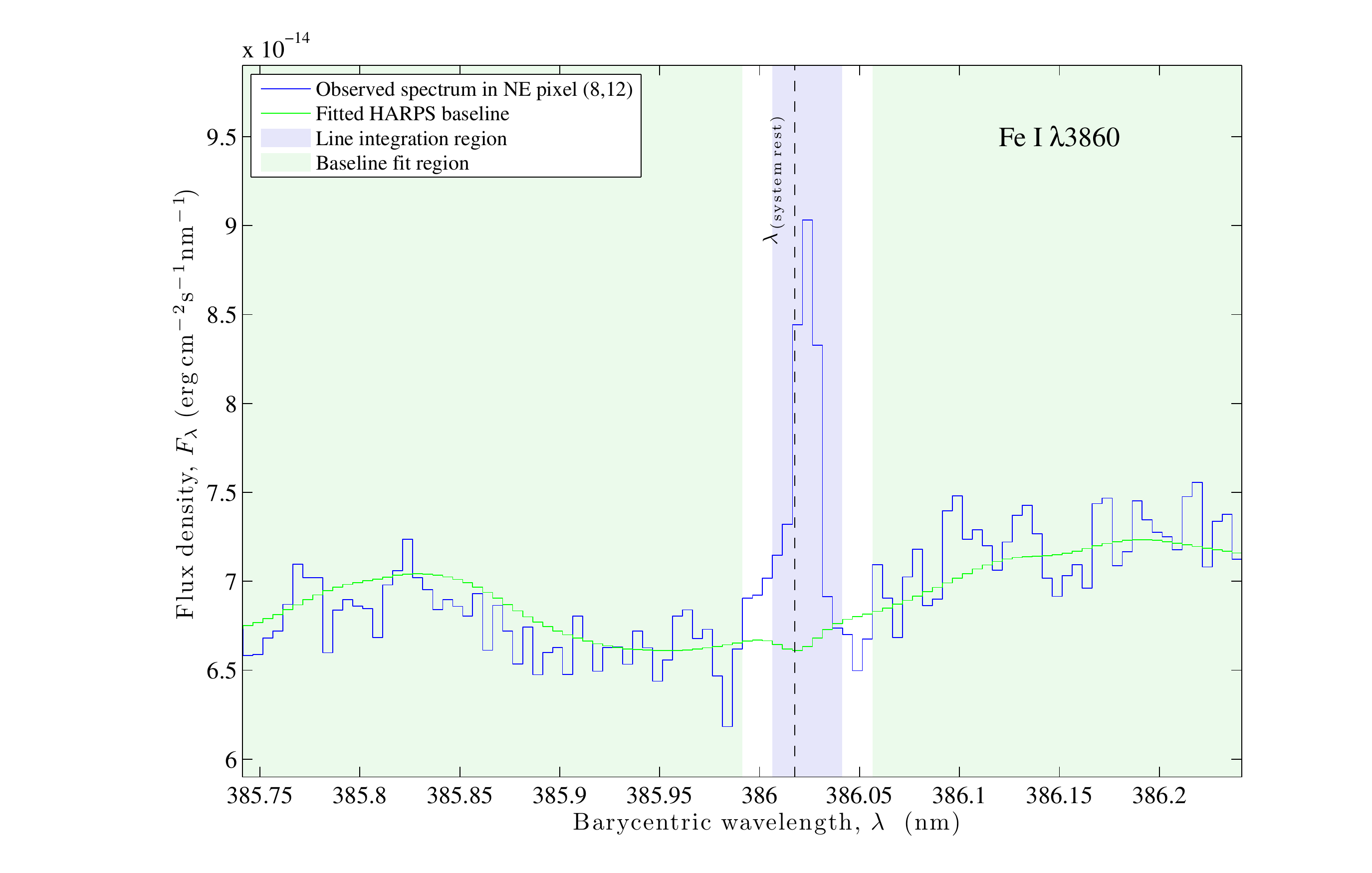}\label{subfiga:accFemidspect}}
\quad
	\subfloat[][]{\includegraphics[trim=35mm 4mm 17mm 0mm, clip=true, width=0.48\textwidth]{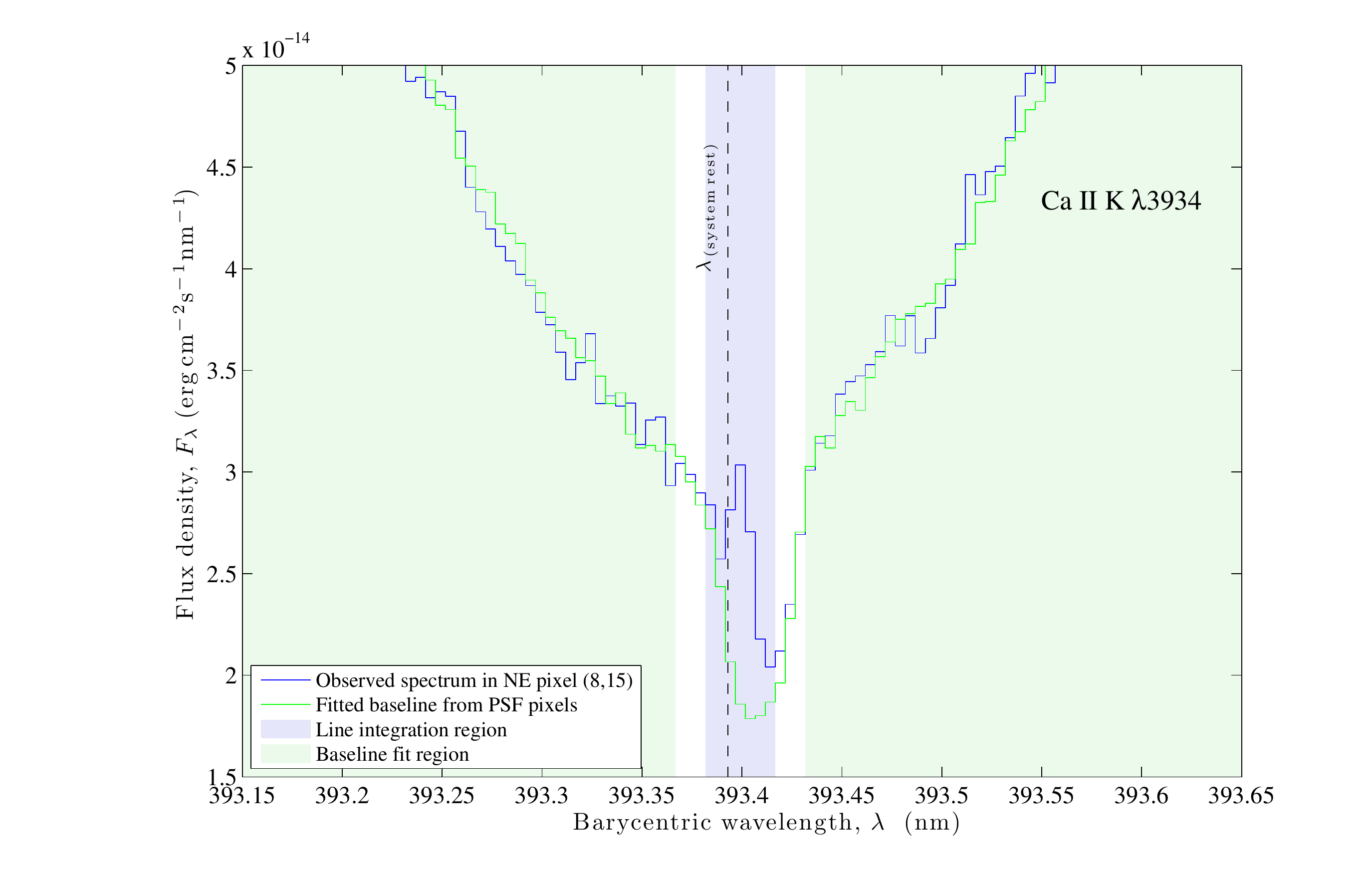}\label{subfigc:accCaHoffspect}}
      \caption{Example spectra of \ion{Fe}{i} $\uplambda$3860 (a) and \ion{Ca}{ii} K (b), both sampled from the NE side of the disk. The regions used for baseline
      fitting and line integration are indicated, as is the systemic rest velocity of $\beta$\,Pic (dashed vertical line).}
      \label{fig:accFemidspect}
\end{figure*}
A spectrum \revone{(with FWHM resolution $\sim$0.004\,nm and sampling of 0.001\,nm/pixel)} of $\beta\,$Pic showing no signs of FEBs, which had previously been obtained from the HARPS spectrograph at the ESO 3.6-m telescope, was convolved with a Gaussian of FWHM=0.0111\,nm (measured average resolution from arc spectra) and rebinned to 0.005\,nm, and used as a reference stellar spectrum, $f_{\star}$. This was then fitted to the observed spectrum, $f_{\textnormal{tot}}$, surrounding each emission line region and in each pixel individually by unconstrained non-linear minimisation of the function
\begin{equation}
f_{\textnormal{disk}}(\lambda) = f_{\textnormal{tot}}(\lambda) - (k_{1}{\lambda}^{2} + k_{2}{\lambda} + k_{3})f_{\star}(\lambda) - k_{4},
\end{equation}
essentially multiplying it with a second-degree polynomial (which was interpolated across the line), but with compensation for offset level. This baseline was then subtracted to recover the gas emission lines. For the inner disk observations of \ion{Ca}{ii} we substituted the HARPS spectrum with a stellar spectrum derived from the mean of eight off-disk pixels with strong stellar PSF contribution, since redshifted \ion{Ca}{ii} components from FEBs were visible in these epochs. The total line flux was found by integrating the spectral energy distribution over the approximated line width at the central wavelength position of the Doppler shifted line (assuming Keplerian rotation of the gas disk). These steps involved some tweaking of parameters (e.g.\ regions for baseline-fit and interpolation, and fitting function) to minimise the effects from noise and certain spectral features. Specifically, the \ion{Ca}{ii} H- and K emission lines are superimposed onto the broad stellar and narrow circumstellar \ion{Ca}{ii} H- and K absorption lines, making the fit of the stellar spectrum very sensitive to the chosen fitting region, and consequently affecting the subtracted baseline. Although this overlap made the reduction more difficult, it also helped by suppressing the background PSF noise of the \ion{Ca}{ii} emission. Fig.~\ref{fig:accFemidspect} shows emission lines of \ion{Fe}{i} and \ion{Ca}{ii} H and K in two selected pixels, together with fitted baselines and regions used in the extraction. The baseline fitting regions extend 100 and 500 wavelength bins on either side of the emission lines for \ion{Fe}{i} and \ion{Ca}{ii}, respectively. The position of the line peaks coincides with expected Doppler-shifts due to Keplerian rotation relative to the systemic velocity of $\beta$\,Pic. Apparent redshifted offsets from systemic velocity in the \ion{Ca}{ii} H- and K \emph{absorption} is a clear indication of the presence of FEBs. 

The signal-to-noise ratio (S/N) of the emission lines was estimated by dividing the baseline-subtracted peak flux with the standard deviation, $\sigma$, of the surrounding spectral region.


\section{Results}\label{sec:res}

\subsection{Disk seen in gas emission}
\begin{figure*}
\centering
	\subfloat[][]{\includegraphics[trim=15mm 4mm 15mm 0mm, clip=true, width=0.49\textwidth]{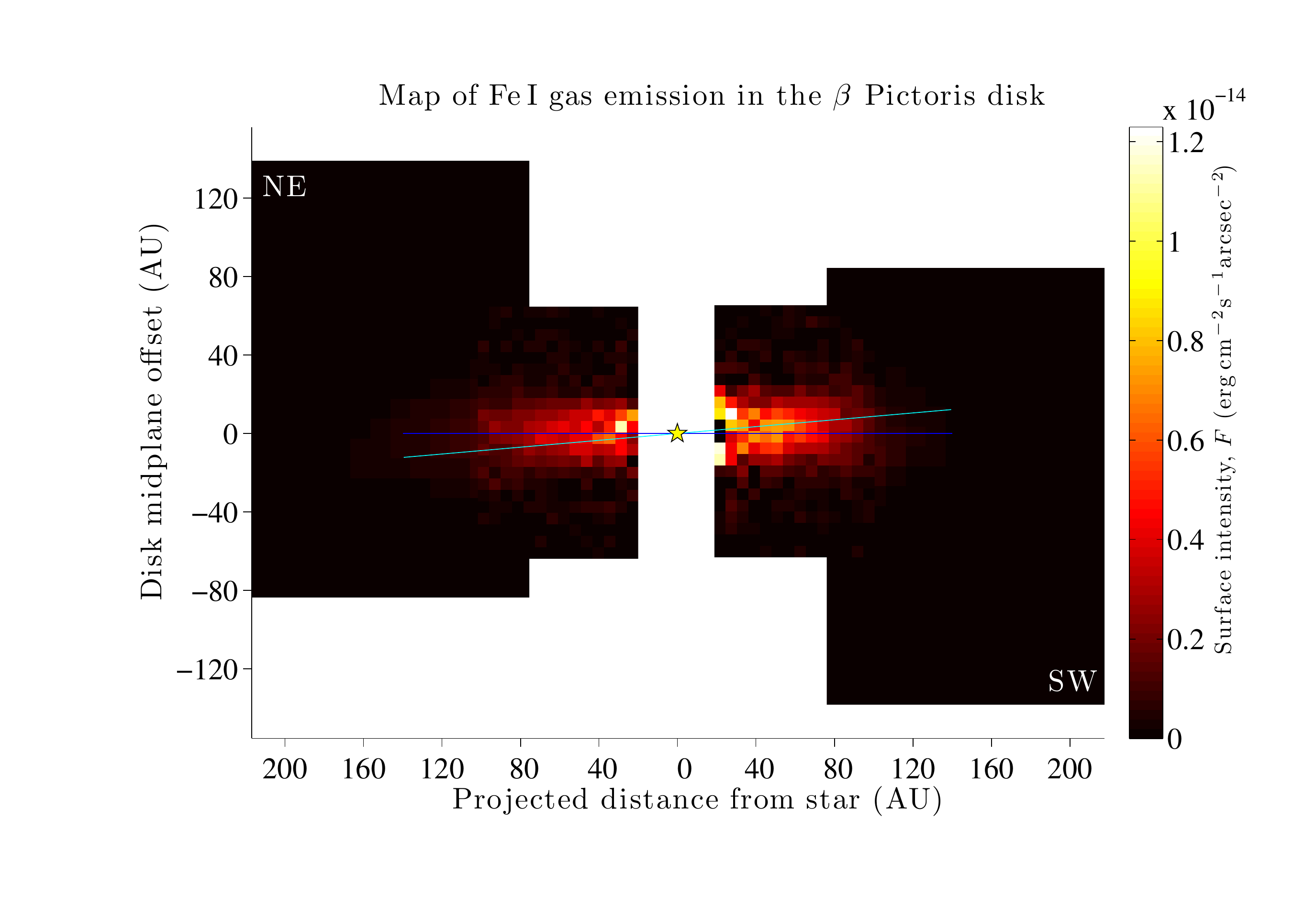}\label{subfiga:Fe_full}}
\quad
	\subfloat[][]{\includegraphics[trim=15mm 4mm 15mm 0mm, clip=true, width=0.49\textwidth]{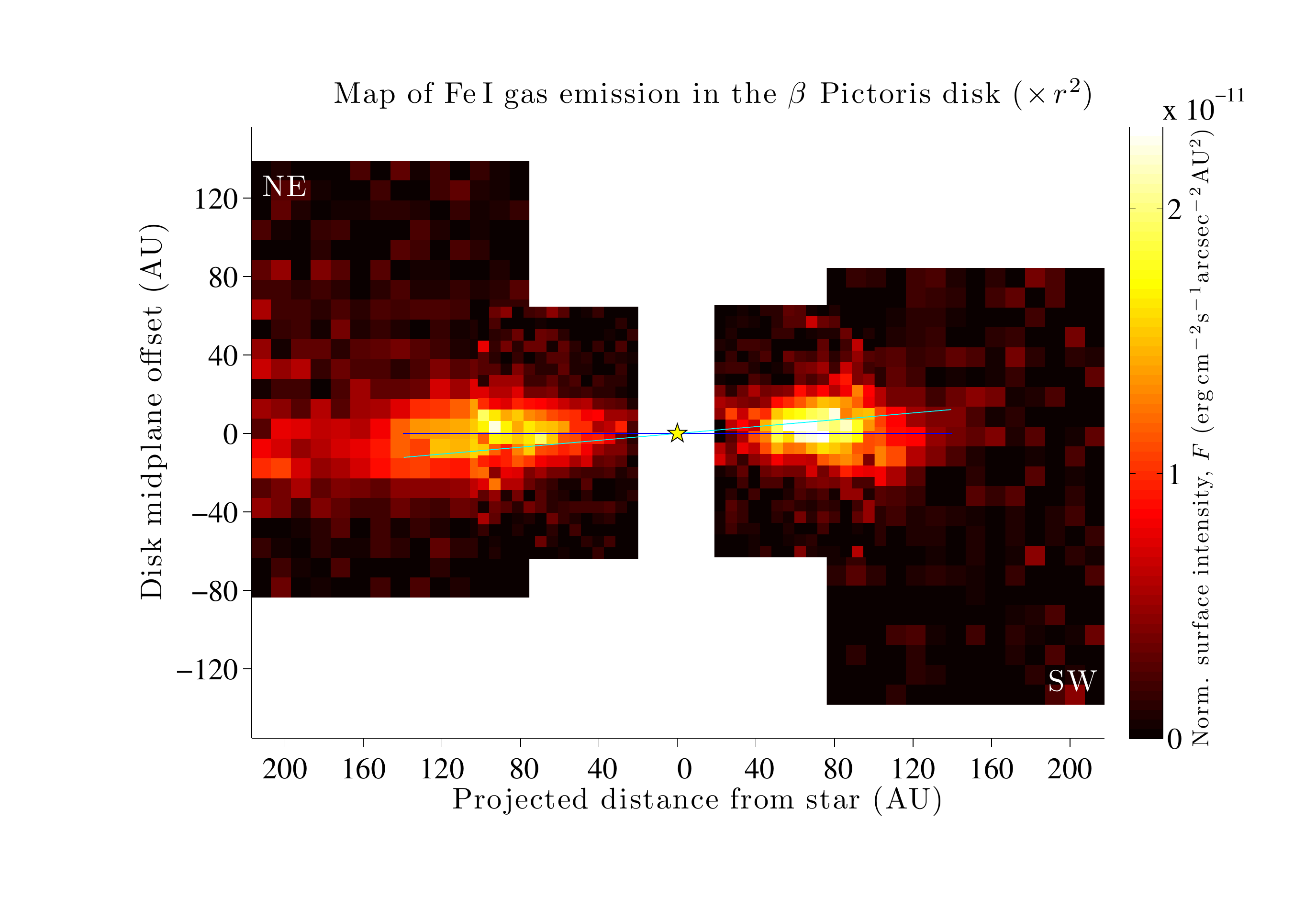}\label{subfigb:Fe_full}}
      \caption{Map of \ion{Fe}{i} emission, multiplied with the projected distance to the star squared in panel (b). The blue line indicates the disk midplane (defined to be at 30.75\degr) while the green line shows the orientation of the inner disk, as found in dust disk observations.}
      \label{fig:Fe_full}
\end{figure*}

\begin{figure*}
\centering
	\subfloat[][]{\includegraphics[trim=15mm 4mm 15mm 0mm, clip=true, width=0.49\textwidth]{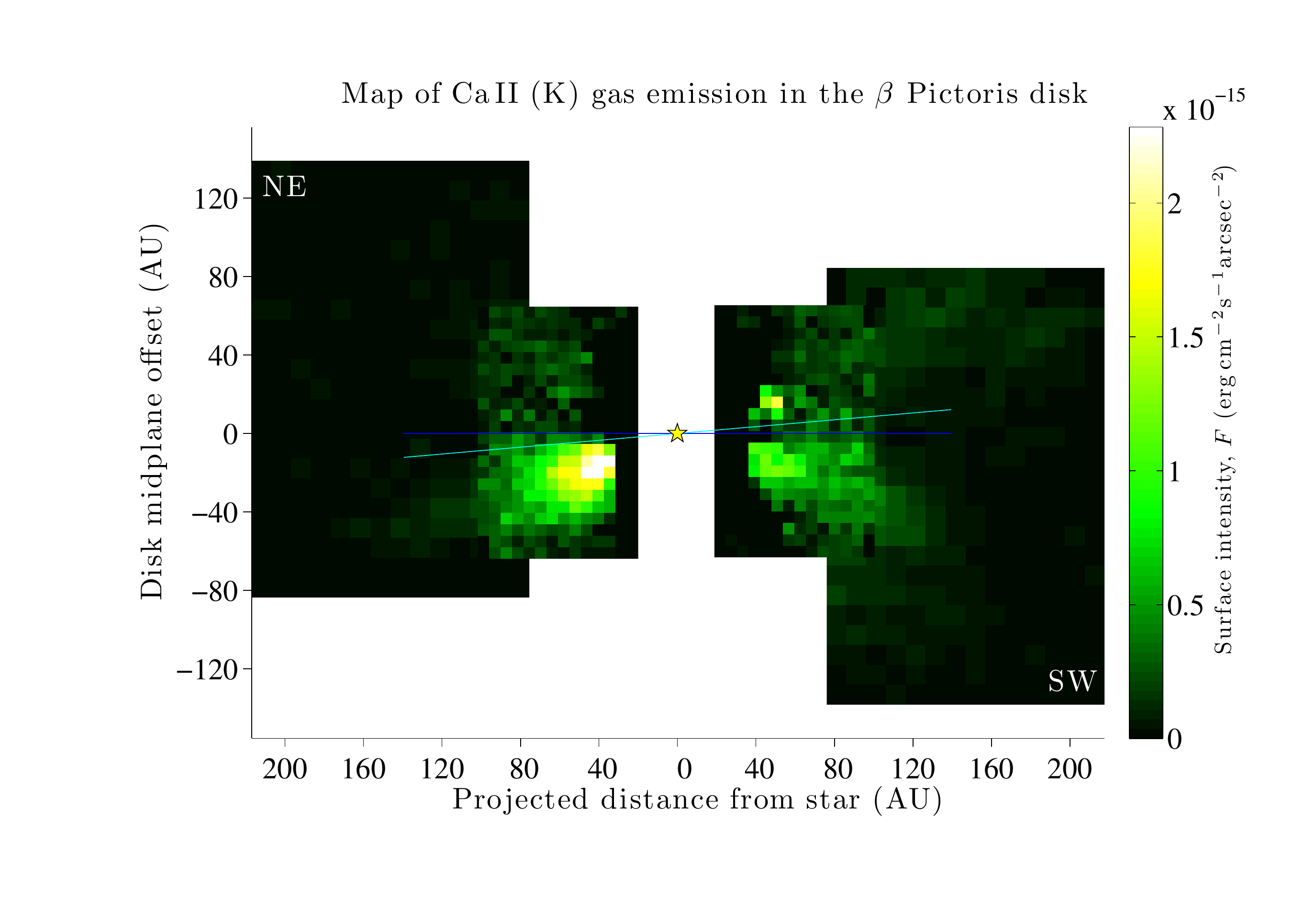}\label{subfiga:Ca_K_full}}
\quad
	\subfloat[][]{\includegraphics[trim=15mm 4mm 15mm 0mm, clip=true, width=0.49\textwidth]{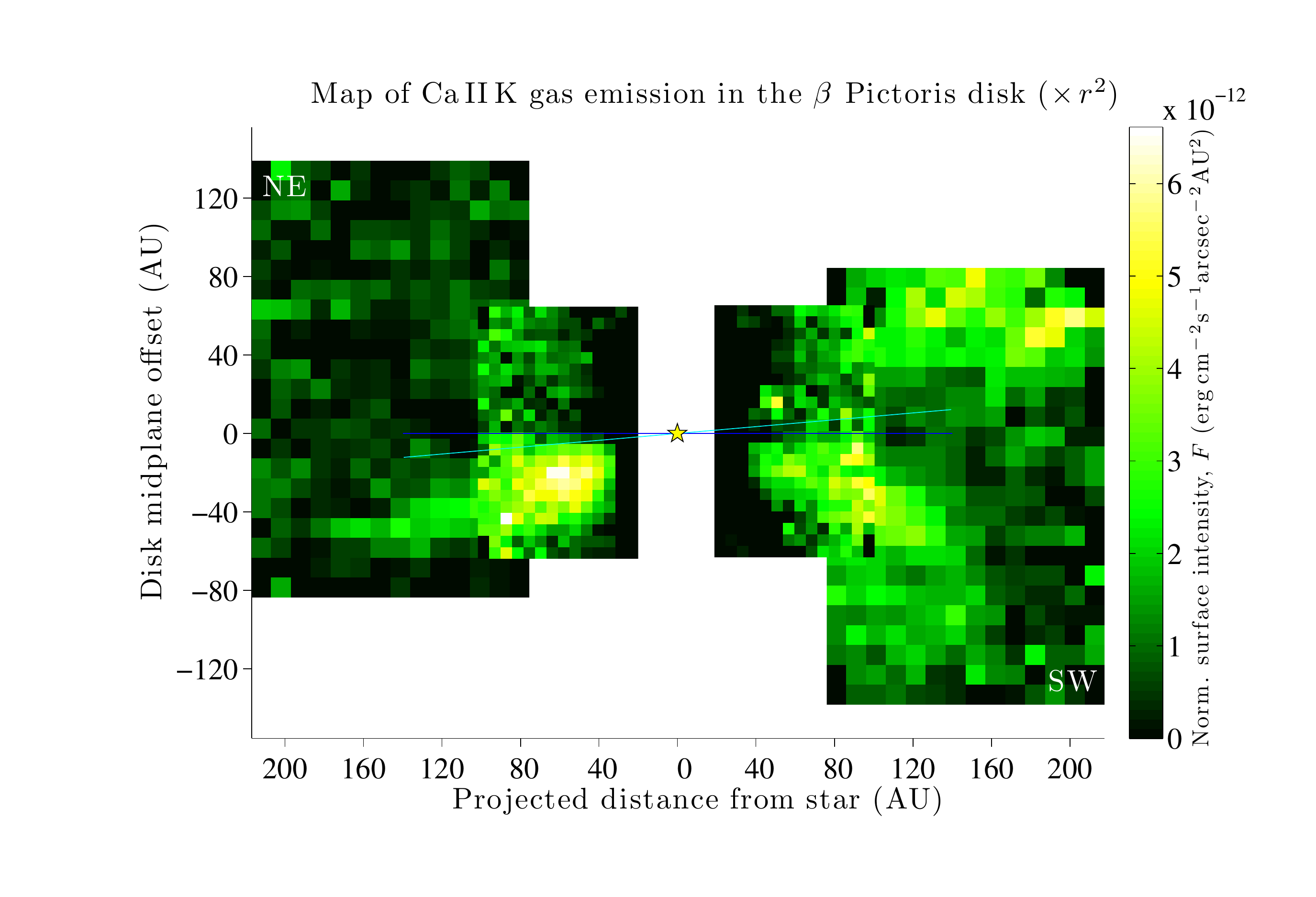}\label{subfigb:Ca_K_full}}
      \caption{As in Fig.~\ref{fig:Fe_full}, but for \ion{Ca}{ii} K.}
      \label{fig:Ca_K_full}
\end{figure*}

\begin{figure*}
\centering
	\subfloat[][]{\includegraphics[trim=15mm 4mm 15mm 0mm, clip=true, width=0.49\textwidth]{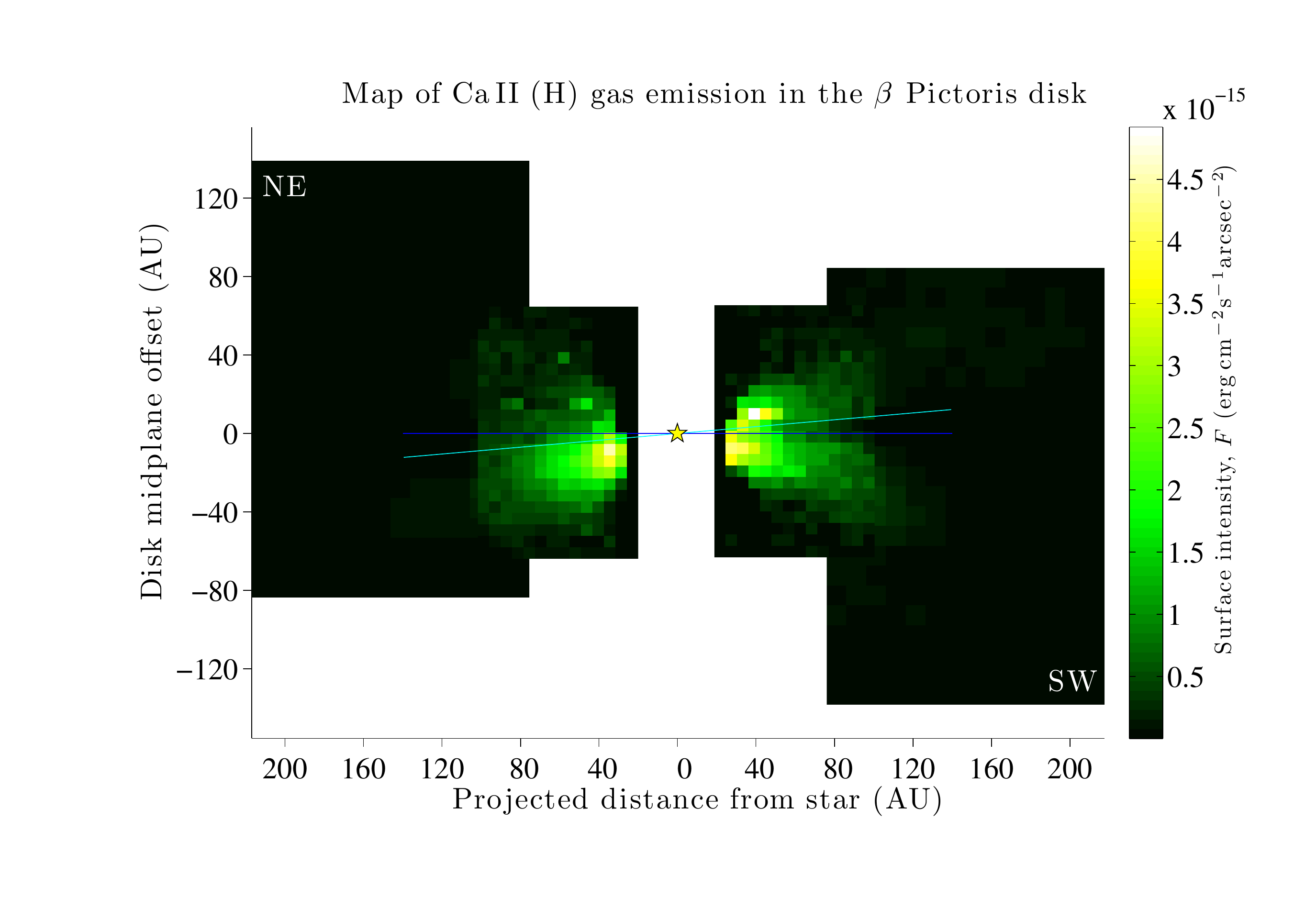}\label{subfiga:Ca_H_full}}
\quad
	\subfloat[][]{\includegraphics[trim=15mm 4mm 15mm 0mm, clip=true, width=0.49\textwidth]{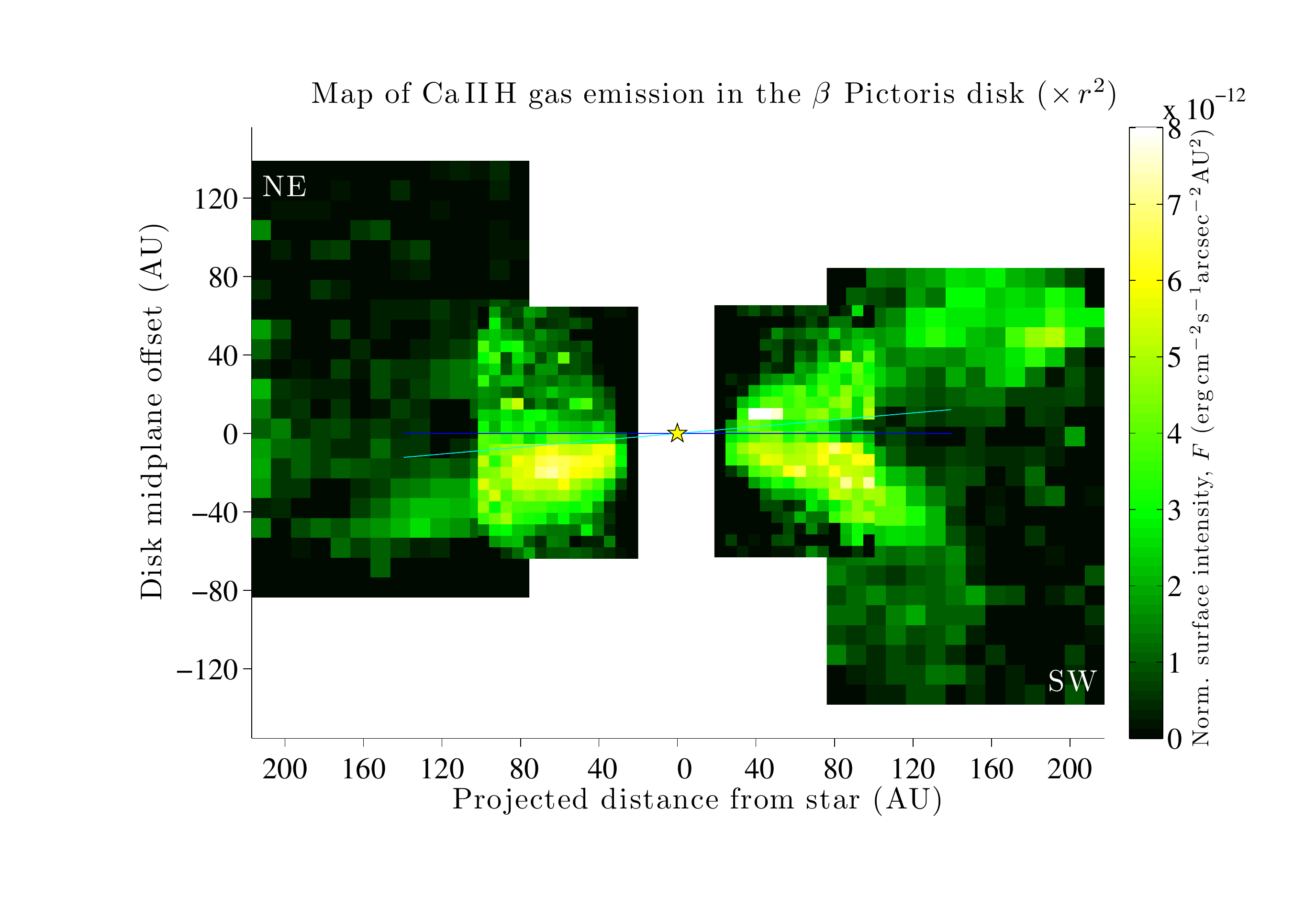}\label{subfigb:Ca_H_full}}
      \caption{As in Fig.~\ref{fig:Fe_full}, but for \ion{Ca}{ii} H.}
      \label{fig:Ca_H_full}
\end{figure*}

\begin{figure*}
\centering
   \includegraphics[trim=7mm 0mm 30mm 0mm, clip=true, width=1.0\textwidth]{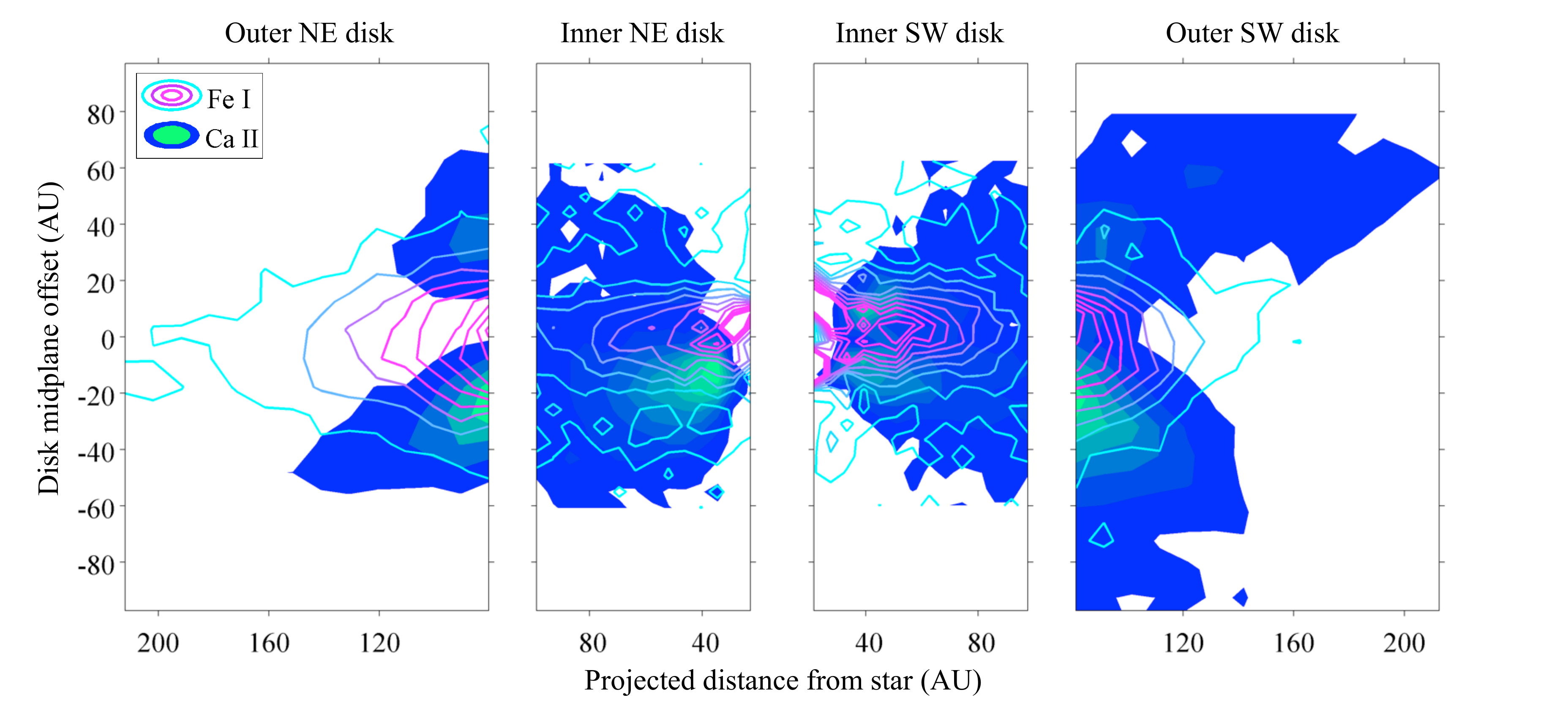}
      \caption{Contour map of \ion{Fe}{i} and \ion{Ca}{ii} H- and K emission. The contour levels of inner and outer panels differ and are just meant to give a clearer picture of the relative positions of dominant \ion{Fe}{i} and \ion{Ca}{ii} emission regions.}
      \label{fig:Fe_and_Ca}
\end{figure*}

Resulting maps of \ion{Fe}{i}\,${\uplambda}3860$ and \ion{Ca}{ii} H- and K emission in the disk of $\beta\,$Pic are presented in Figs.~\ref{fig:Fe_full}--\ref{fig:Ca_H_full}, where the (b) panels display the intensity scaled with projected distance to the star squared to bring out detail in the outer regions. The optical dust-disk midplane is marked by a blue line, and the tilted $4.5\degr$ secondary disk with a cyan-coloured line. \ion{Fe}{i}$\,{\uplambda}3860$ emission is clearly visible along the disk midplane, while \ion{Ca}{ii} H- and K emission is strongest in higher latitude areas on the NE and SW side. Both exhibit a highly asymmetric distribution; \ion{Fe}{i} being fainter but more extended in the NE than in the SW, and \ion{Ca}{ii} showing the opposite behaviour, in addition to asymmetries relative to the disk midplane. \ion{Fe}{i} is detected at disk radii out to the 210\,AU limit of our observations in the NE, while \ion{Ca}{ii} appears to be reaching beyond the observed $\sim$100\,AU latitudes on the SW side. Data from the innermost 2--3 columns ($\sim$20--30\,AU) are very noisy due to strong contamination by light from the stellar PSF, therefore it is hard to tell if there is a decline in flux inward of the peaks at $\sim$40\,AU. However, inspection of the maps normalised with distance squared suggests that there is a peak in the \emph{column density} of \ion{Fe}{i} at $\sim$90\,AU in the NE and $\sim$70\,AU in the SW. Because \ion{Ca}{ii} H- and K emission is optically thick, we cannot immediately draw similar conclusions for \ion{Ca}{ii}, as discussed in detail in Sect.~\ref{subsec:caiimodel}. 

A brightness difference can be observed in the overlapping regions between observations of the outer and inner disks. This can be caused by a number of factors and is discussed in more detail in Appendix~\ref{sec:calibcomp}. 

Fig.~\ref{fig:Fe_and_Ca} clearly shows how the emission contours from neutral Fe outlines the disk midplane, while singly ionised Ca emission appears weaker in that same region, instead dominating at high latitudes and increasing with radial distance, forming a characteristic butterfly shape. Note that this figure is only presented to highlight the relative positions of regions with dominating \ion{Fe}{i} and \ion{Ca}{ii} gas emission, and that the contour levels of the inner and outer parts differ.


\section{Analysis}\label{sec:ana}

\revone{Our primary} interest is to find the spatial gas density distribution in the disk. The location of the gas gives clues to its origin, and can also put constraints on potential planet-disk interaction (as has indeed been done for the observed spatial \emph{dust} distribution). Of particular interest is the \ion{Ca}{ii} observed at surprisingly high scale-heights above the disk midplane, and the possible mechanisms that may have placed the Ca there. 

The problem of inverting observed flux into gas density is not trivial, because we are observing an essentially edge-on projection of the disk emission, where the emission at each projected distance is the contribution from many locations in the disk along the line of sight. If we furthermore consider the possibility of an asymmetric disk, the problem of finding the gas density distribution from the observed projected flux becomes degenerate. To break the degeneracy, we assumed that the disk is cylindrically symmetric. Since we observe significant differences between the NE and SW sides of the disk, we know that this assumption is of limited validity; we therefore made independent inversions of the two sides to get an impression of the uncertainties involved. To simplify the procedure even more, we divided the problem into two steps:

\begin{enumerate}
\item De-project the \ion{Fe}{i} $\uplambda$3860 emission from the disk to obtain the spatial luminosity density as a function of cylindrical radius $r$ and height $z$ (assuming cylindrical symmetry).

\item Use the derived 2D luminosity density profile to derive the Fe gas density, with the help of the ionisation, thermal balance, and level population statistical equilibrium code \textsc{ontario}.
\end{enumerate}

The inversion from the \ion{Fe}{i} $\uplambda$3860 luminosity density to the number density of Fe is complicated by the complex energy level structure of \ion{Fe}{i}. The level population is dominated by photo-excitation, hence the need for a non-LTE (local thermal equilibrium) solver (part of \textsc{ontario}). \revone{Furthermore, the \ion{Fe}{i} $\uplambda$3860 emission is expected to be optically thin; for a line width of 1.5\,km\,s$^{-1}$ \citep{Crawford1994}, the mid-plane column density of \ion{Fe}{i} towards the star is 7$\times$10$^{11}$\,cm$^{-2}$ \citep{Zagorovsky2010}, and conservatively assuming \textit{all} atoms to be in the ground state, the optical depth of the $\uplambda$3860 transition becomes $\tau \sim 0.1$.}

\revone{In contrast, assuming a solar abundance ratio of Fe/Ca \citep{Zagorovsky2010}, the \ion{Ca}{ii} H- and K lines are strongly optically thick ($\tau \sim 100$ in the mid plane). We therefore did not attempt to use the \ion{Ca}{ii} lines to derive the gas density of Ca, since even small deviations from cylindrical symmetry in density will give large variations in emitted flux, making the inversion strongly degenerate.} Instead, we used the spatial gas distribution derived from \ion{Fe}{i}, solved for the radiative transfer and computed the expected \ion{Ca}{ii} H- and K emission from the disk, in an attempt to check whether there is chemical segregation, as has been previously suggested \citep{Beust2007}.

In the following subsections, the inversion steps are outlined in detail.

\subsection{Finding the \ion{Fe}{i} $\uplambda$3860 luminosity density profile}\label{subsec:fit}
\begin{figure*}
\centering
   \includegraphics[trim=10mm 43mm 10mm 30mm, clip=true, width=1.0\textwidth]{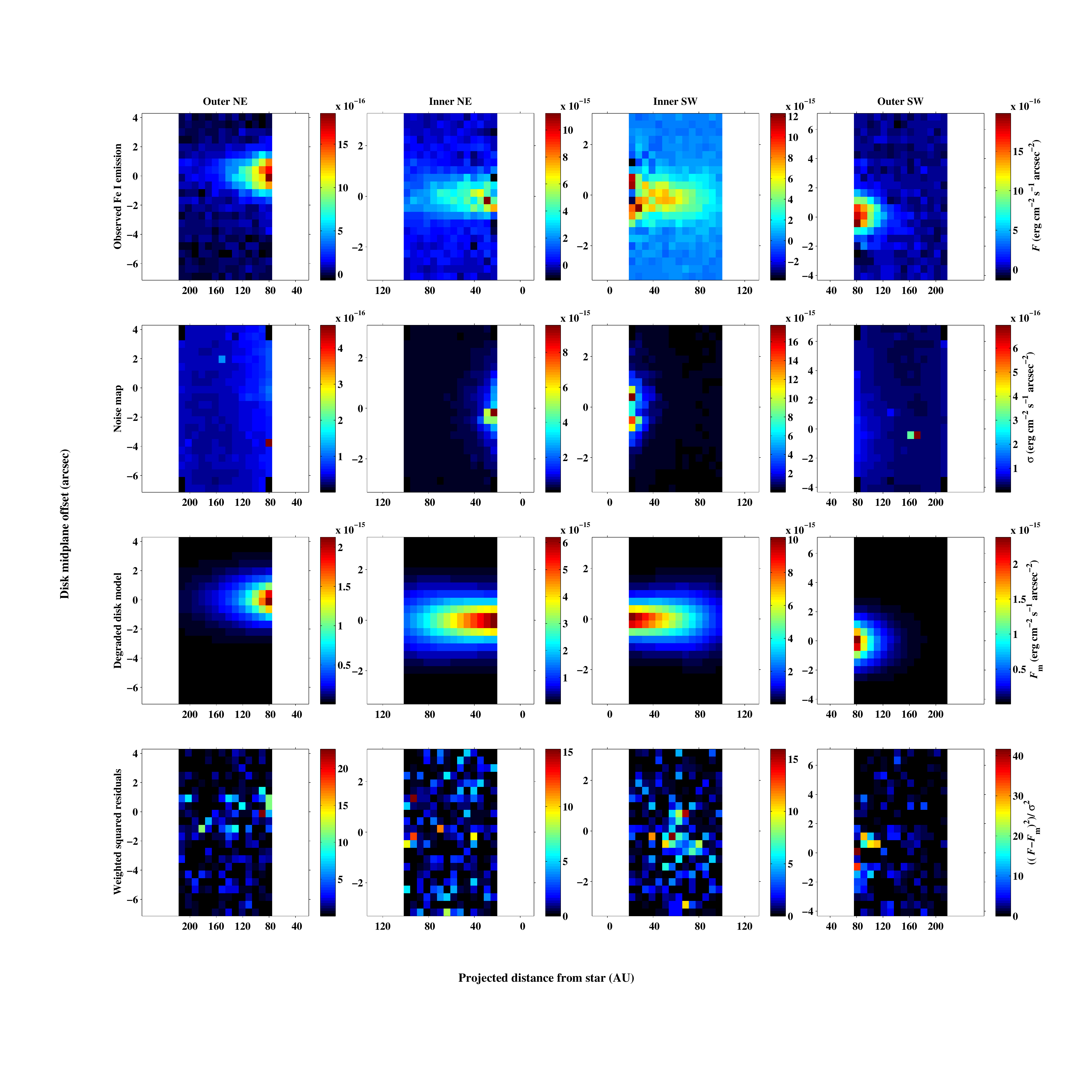}
      \caption{Display of input data and results from ${\chi}^{2}$-minimisation disk-modelling of the \ion{Fe}{i} emission. The rows of panels show, from the top down: observed \ion{Fe}{i} emission maps, derived noise maps, degraded model maps, and maps of the weighted squared residuals. The reduced ${\chi}^{2}$ values are 1.9 (NE) and 2.3 (SW).}
      \label{fig:Fe_model_subs}
\end{figure*}
To find the de-projected luminosity density profile of \ion{Fe}{i} $\uplambda$3860, we assumed cylindrical symmetry and fitted a few-parametric functional form to the observed flux distribution. In line with \citet{Brandeker2004}, we used a broken power-law for the radial dependence but introduced two new parameters for the height dependence ($\gamma$) and the scale-height dependence ($\delta$) on the midplane radius $r$:
\begin{equation}\label{lum}
L(r,z)=l_{0}\left[\frac{2}{(r/r_{0})^{2\alpha}+(r/r_{0})^{2\beta}}\right]^{1/2}\exp\left[{-\left(\frac{z}{h(r)}\right)^{\gamma}}\right],
\end{equation}
where $\alpha$ and $\beta$ are the parameters describing the shape of the power-law on either side of the inflection point $r_{0}$, and the vertical scale-height at each $r$ defined as
\begin{equation}\label{scaleheight}
h(r)=h_{0}\left(\frac{r}{r_{0}}\right)^{\delta}.
\end{equation}
The parameter $\delta$ describes the `flaring' of the disk. For $\delta = 0$ we have a constant thickness disk, while for $\delta = 1$ the disk scale height increases linearly with radius. For $\delta > 1$ we have a flaring disk, i.e.\ the disk opening angle increases with radius. The parameter $\gamma$ determines how quickly the density drops with height. For a gas disk in thermostatic equilibrium we would expect the number density to drop as a Gaussian function, i.e.\ $\gamma = 2$. The dust distribution, on the other hand, is best fitted by closer to an exponential ($\gamma \approx 0.5$; Ahmic et al.\ 2009).

For a given set of parameters, $L(r,z)$ was projected onto a plane from edge-on to a resolution ten times better than the observations. The projected flux distribution was then convolved with the seeing recorded for the observation by using a Gaussian of the same FWHM, and re-sampled to the grid of the observations for comparison with observed data. By computing the residuals weighted by the noise, the best-fit values for the parameters $l_{0}$, $h_{0}$, $r_{0}$, $\alpha$, $\beta$, $\gamma$, and $\delta$ were found by $\chi^2$-minimisation. Fig.~\ref{fig:Fe_model_subs} displays maps of observed \ion{Fe}{i} emission, estimated noise, best-fit degraded model, and weighted squared residuals of the model fit. Resulting parameter values from fitting of the gas-disk profile and associated errors are presented in Table~\ref{table:params}. The errors were estimated using the Monte-Carlo method to make repeated fits to the data with added noise drawn from the calculated noise distribution in each pixel.

\begin{table}
\caption{Best-fit parameters and errors.}             
\label{table:params}      
\centering          
\begin{tabular}{ l r r r r r r }     
\hline\hline 
& \multicolumn{4}{c}{\ion{Fe}{i} $\uplambda$3860 luminosity} & \multicolumn{2}{c}{Density} \\      
& \multicolumn{2}{c}{NE} & \multicolumn{2}{c}{SW} & NE & SW \\
Parameter & value & $\sigma$ & value & $\sigma$  & value & value \\ 
\hline 
   $l_{0}$ ($10^{-19}$\,erg\,s$^{-1}$cm$^{-3}$) & 5.54 & 1.10 & 8.59 & 1.60 & --- & --- \\  
   $n_{0}$ ($10^{3}$\,cm$^{-3}$) &  --- & ---  & ---  & ---  & 1.19 & 1.76\\  
   $h_{0}$ (AU) & 16.4 & 1.0 & 16.9 & 1.0 & 21.8 & 20.4\\ 
   $r_{0}$ (AU) & 87.7 & 4.9 & 85.4 & 3.3 & 94.3 & 83.9\\ 
   $\alpha$ & 4.98 & 0.15 & 6.88 & 0.24 & 3.00 & 3.86\\  
   $\beta$ & 1.15 & 0.28 & 1.38 & 0.29  & 0.83 & 0.93\\ 
   $\gamma$ & 1.50 & 0.11 & 1.57 & 0.11  & 1.26 & 1.33\\ 
   $\delta$ &  0.93 & 0.07 & 1.19 & 0.10  & 0.96 & 1.21\\ 
\hline                  
\end{tabular}
\end{table}

\subsection{Finding the gas density from the luminosity density}\label{subsec:ont}

To find out how much gas we need to produce the inferred luminosity density of a \ion{Fe}{i} $\uplambda$3860, we have to answer these questions:
\begin{enumerate}
\item What is the number fraction of Fe in the gas?
\item What is the fraction of Fe in the neutral, atomic state?
\item What is the fraction of the \ion{Fe}{i} that is radiating at $\uplambda$3860?
\end{enumerate}
Because $\beta\,$Pic is an A-type star, we expect its radiation field to dissociate almost all molecules in its disk, leaving only atomic and ionised gas components. This simplifies the estimate of free Fe atoms because we can ignore chemical interactions, and only need to consider the ionisation state. For this we used the \textsc{ontario} code \citep{Zagorovsky2010}, which self-consistently models the ionisation states and gas temperatures in debris disks around A- and F-type stars given a known stellar flux and gas/dust profile. It performs a full statistical equilibrium computation for several atomic species to find dominant absorption and emission lines. The local gas temperature is determined by the balance of heating and cooling mechanisms, and can, for the assumed optically thin gas, be treated in independent computation bins. The dominating heating processes for $\beta\,$Pic are photoelectric heating by dust, and photoionisation of gas by stellar UV light, while cooling is dominated by fine-structure transitions (in particular from \ion{C}{ii} and \ion{O}{i}, as recently detected in emission by the \textit{Herschel Space Observatory}, Brandeker et al.\ 2012). A detailed description of the code can be found in \citet{Zagorovsky2010}.

\begin{figure*}
\centering
	\subfloat[][]{\includegraphics[width=0.45\textwidth]{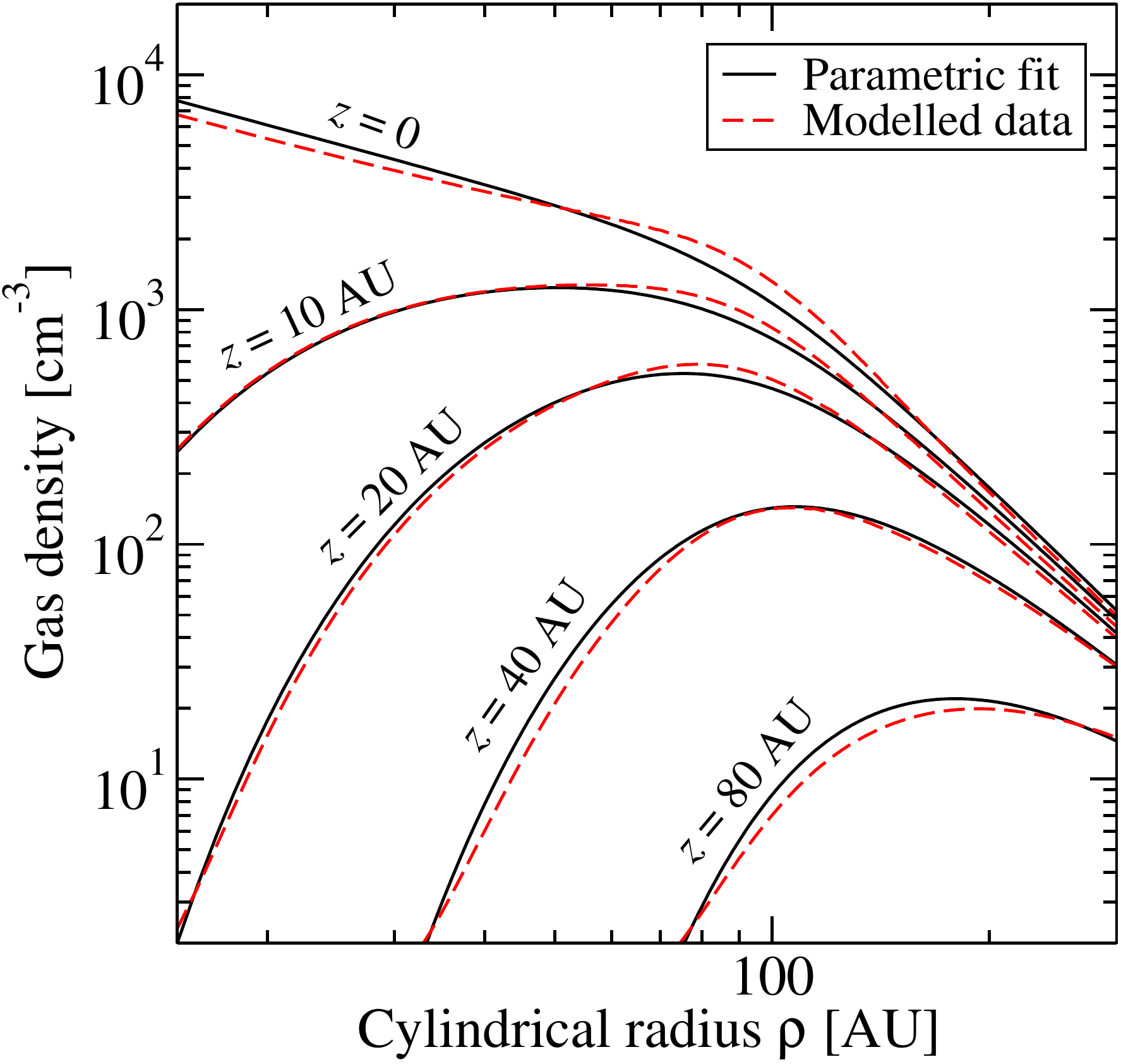}\label{subfiga:Fe_ontario_NEz}}\subfloat[][]{\includegraphics[width=0.45\textwidth]{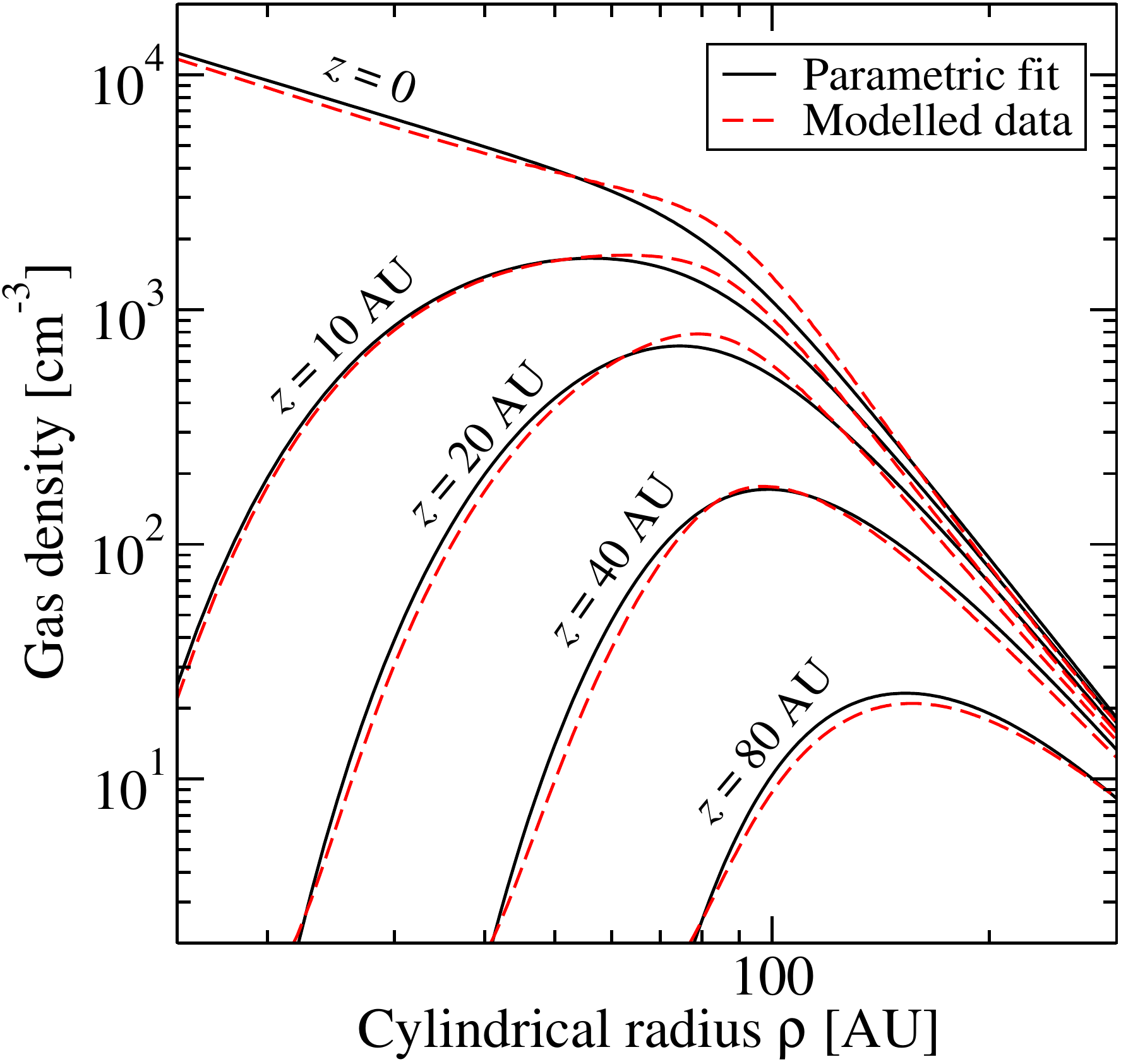}\label{subfiga:Fe_ontario_SWz}}
	
	\subfloat[][]{\includegraphics[width=0.45\textwidth]{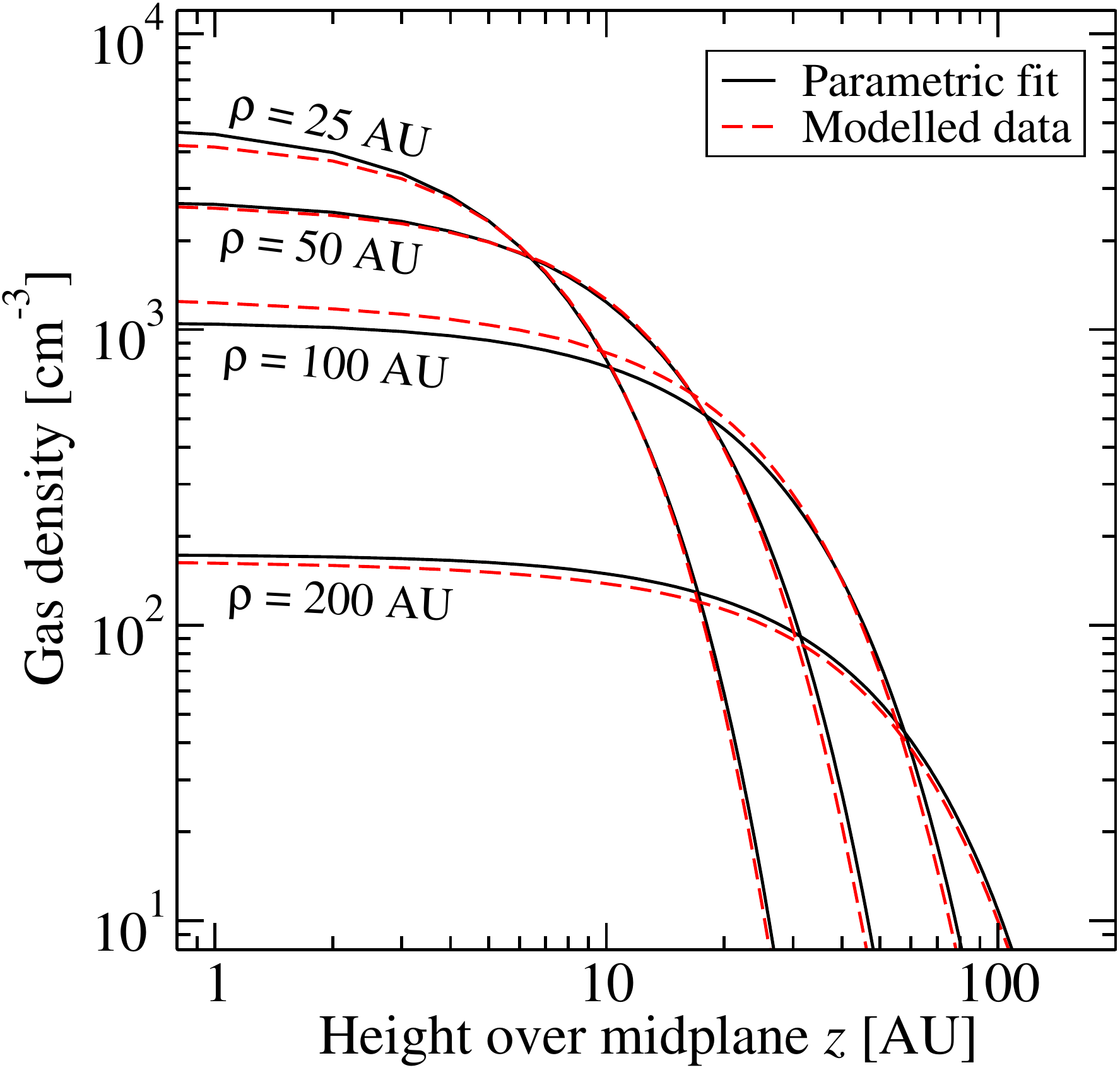}\label{subfiga:Fe_ontario_NEr}}\subfloat[][]{\includegraphics[width=0.45\textwidth]{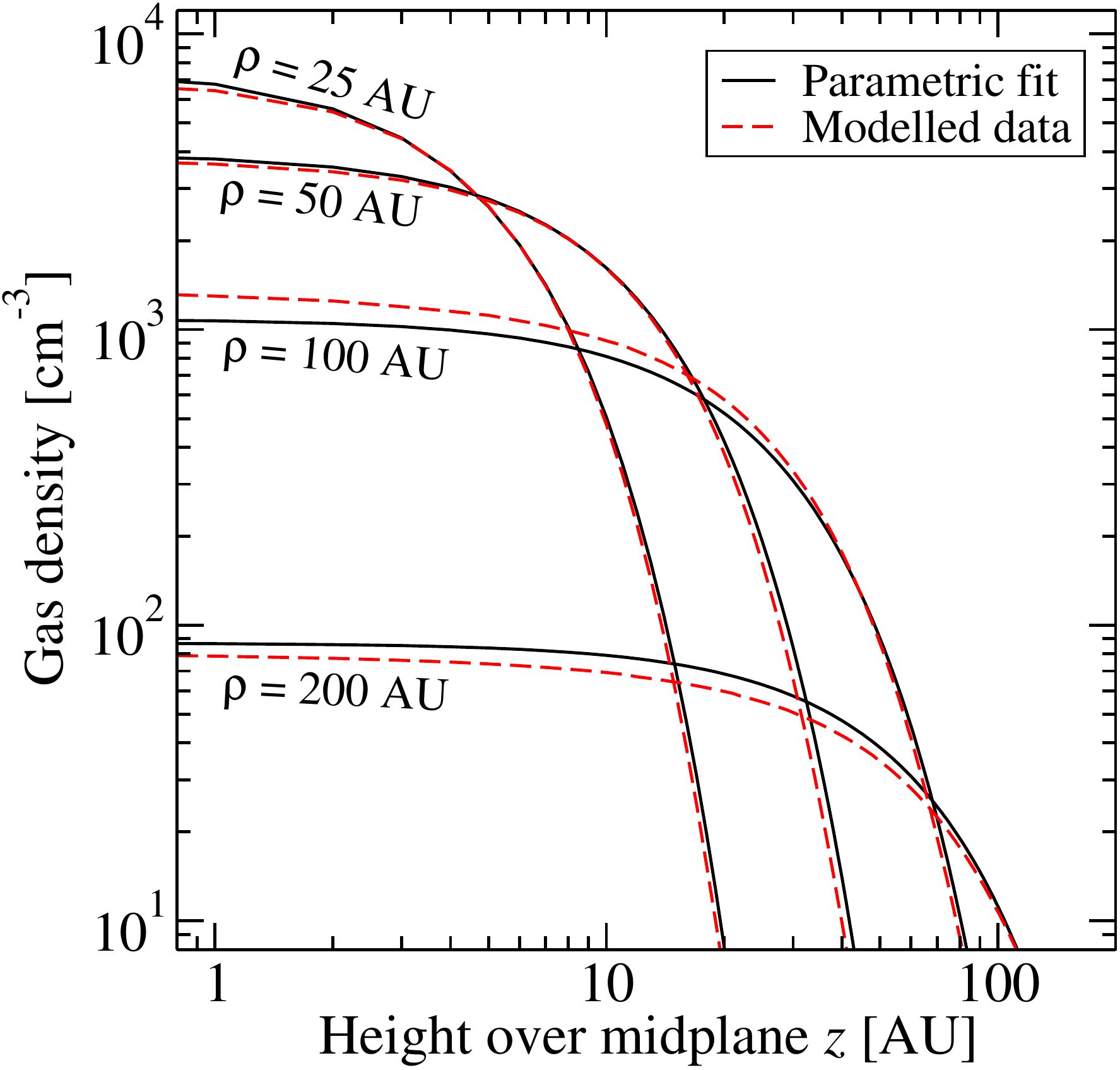}\label{subfiga:Fe_ontario_SWr}}
	
      \caption{Hydrogen-equivalent gas density inferred from \ion{Fe}{i} $\uplambda$3860 with parametric fits. Multiply with used abundance fraction to obtain the density of any specific element (well-mixed gas is assumed). NE to the left and SW to the right; top panels show cuts parallel to the midplane at heights $z$ above the plane and the bottom panels cuts orthogonal to the midplane at cylindrical radii $r$.}
      \label{fig:ontario_invFe}
\end{figure*}

By iterating \textsc{ontario} with different gas densities, the density that best reproduced the given emission for each grid cell was found. The abundance  of elements in the gas was assumed to be solar \citep{Grevesse1998}, except for H (0.1\% of solar), He (none), and C and O (400 times solar) as  inferred from observations \citep{Brandeker2012}. The resulting gas profiles (separately for the NE and SW sides) are presented in Fig.~\ref{fig:ontario_invFe} together with a best analytical fit, using the same parameters as $L(r,z)$ of Eq.~\ref{lum} but with the luminosity density $l_0$ replaced by a number density $n_0$, which is a parameter that refers to the number density of H had it been at solar abundance. The number density of any given element can then be found by multiplying $n_0$ with the abundance relative to H (had H been at solar abundance). For instance, the number density of Fe is given by $n_{\mathrm{Fe}} = 2.8\times10^{-5} n_0$ and C by $n_{\mathrm{C}} = 9.8\times10^{-2} n_0$.

\section{Discussion}

Looking at the spatial distribution, we confirm the strong asymmetry between the NE and SW sides of the disk, where the SW gas density seemingly drops much quicker with radius at a rate $\propto r^{-3.9}$ compared to the NE side $\propto r^{-3.0}$. Neither side shows evidence for flaring, as $\delta \sim 1$. This means that the $h/r$ ratio is fairly constant and close to $h_0/r_0 \sim 0.2$. The height dependence with $\gamma \sim 1.3$ is closer to exponential ($\gamma = 1$) than the Gaussian ($\gamma = 2.0$) one would expect from hydrostatic equilibrium. This could be an indication that the gas dynamics is significantly affected by the dust grains, as found by \citet{Fernandez2006} to be expected for an ionised gas interacting with charged dust.

\subsection{The \ion{Ca}{ii} profile}\label{subsec:caiimodel}
In contrast to the detected \ion{Fe}{i} emission (Fig.~\ref{fig:Fe_full}), the \ion{Ca}{ii} H- and K emission (Figs.~\ref{fig:Ca_K_full}--\ref{fig:Ca_H_full}) is observed far away from the midplane. An interesting question is whether there is a mechanism preferentially transporting Ca to far heights above the midplane \citep[as argued by e.g.][]{Beust2007}, or if there is another reason for \ion{Ca}{ii} emission to be preferentially emitted there. A significant difference between \ion{Ca}{ii} and \ion{Fe}{i} \citep[and the other spatially mapped species \ion{Na}{i} from][]{Brandeker2004} is that, since the metals are all strongly ionised, the number density of \ion{Ca}{ii} is 15--500 times higher than the neutral \ion{Fe}{i} (and \ion{Na}{i}). Furthermore, the H- and K transition probabilities of \ion{Ca}{ii} are 10$\times$ higher. Together, this means that while the neutral transitions are optically thin, the \ion{Ca}{ii} lines are optically thick. A reason for the paucity of midplane emission could therefore be that the star is obscured by the gas, and that the only photons being scattered from the H- and K transitions in the midplane either come from secondary sources (e.g.\ photons scattered down from higher altitudes) or from far out in the wings, where the lines are still optically thin.

\begin{figure}
\centering
   \includegraphics[width=0.4\textwidth]{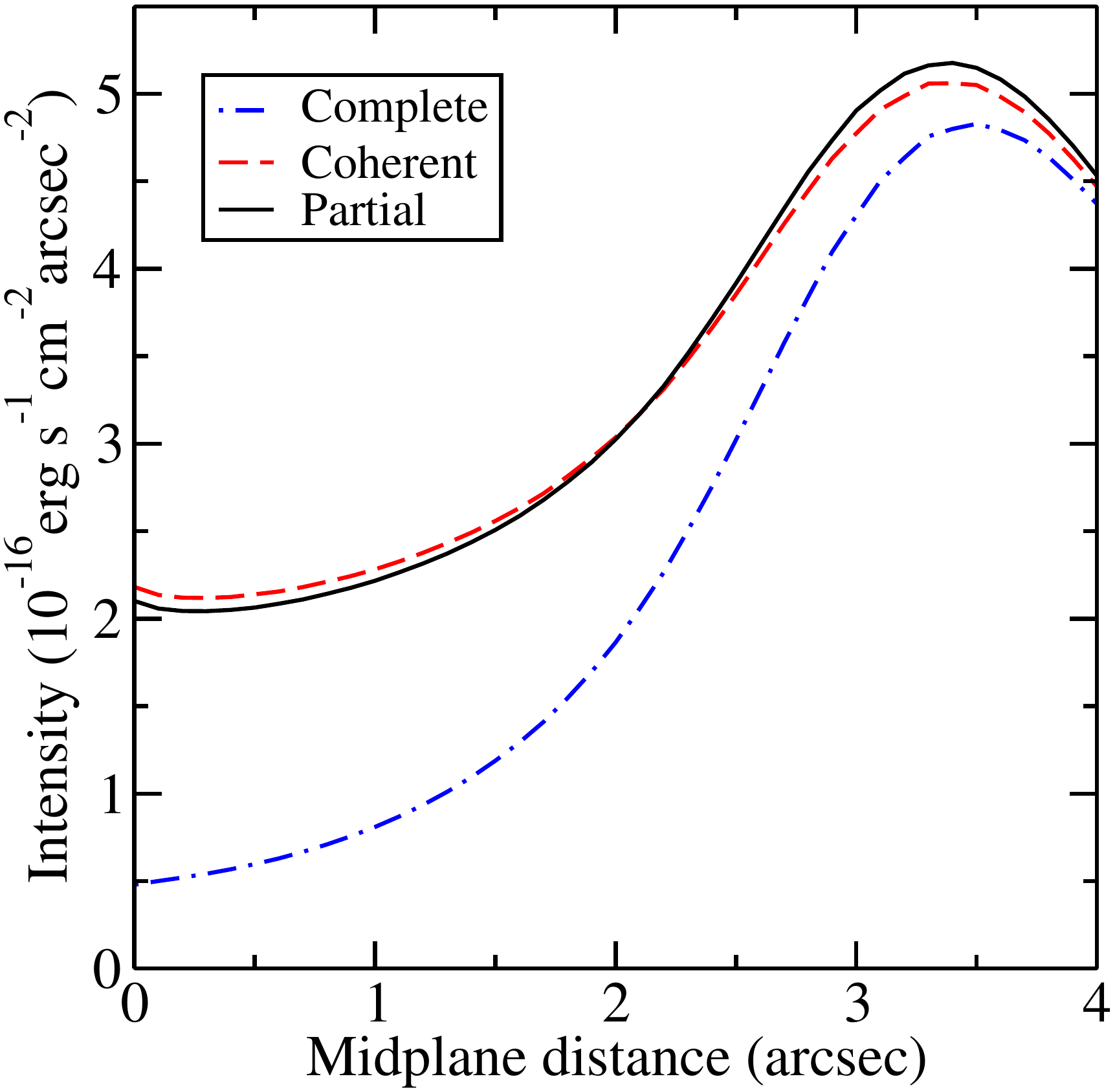}
      \caption{modelled \ion{Ca}{ii} H emission profile, using three different photon scattering redistribution functions. The figure shows a vertical cut at a distance
      of 120\,AU at the SW side.}
      \label{fig:redist}
\end{figure}

To test the idea that the optical properties are responsible for the difference in appearance between the \ion{Fe}{i} $\uplambda$3860 and \ion{Ca}{ii} H- and K emission maps, we implemented a simple radiative transfer code to compute the emerging emission map. Because we expect the lines to be optically thick, the line shape is important as the optical depth will vary over the line profile. Furthermore, because the gas is orbiting the star, there will be a velocity field that will Doppler-shift the line profile depending on the velocity projected on the line of sight. In detail, our model is based on the following assumptions and parameters:
\begin{enumerate}
\item The spatial distribution of gas is assumed to be the one in Table~\ref{table:params}, inferred from the \ion{Fe}{i} $\uplambda$3860 emission (we assume different profiles for the NE and SW sides). The number density of \ion{Ca}{ii} is then computed using \textsc{ontario} with the same parameters as used when inverting the \ion{Fe}{i} emission profile .
\item For the velocity field, circular Keplerian orbits are assumed with a dynamical mass equal to $M_{\mathrm{dyn}} = 1.4\,M_{\odot}$ \citep{Olofsson2001}.
\item As an intrinsic line profile, a Voigt profile is assumed with broadening parameter $b = 1.5$\,km\,s$^{-1}$. This is close to, but slightly below, the broadening parameter measured in absorption \citep[$b = 2.0\pm0.7$\,km\,s$^{-1}$;][]{Crawford1994}, which on the other hand probes the full column of gas along the midplane.
\item A 2D luminosity profile is computed, where the emission as a function of wavelength is derived as a function of cylindrical radius and height above the midplane. The luminosity profile Doppler-shifted by the projected velocity is then integrated through the disk for the line of sight, taking the (also Doppler-shifted) line opacity along the line of sight into account.
\item For the line scattering redistribution function we try three different cases:
\begin{enumerate}
	\item Complete redistribution, that is, the outgoing photon is re-emitted at a wavelength with a probability according to the (Doppler-shifted) line profile, independent of the wavelength of the absorbed photon. This is a good approximation if the \ion{Ca}{ii} is disturbed (by a collision) before emitting again, which is likely not the case in the present circumstances because of the strength of the transition and the low density of gas.
	\item  Coherent scattering, that is, the re-emitted photon has exactly the same wavelength as the absorbed one (only Doppler-shifted). This is likely a better approximation, but essentially assumes that there is no Doppler-shift of the photons due to thermal motion, only the Keplerian velocity field.
	\item Partially coherent redistribution, which assumes coherent scattering \textit{in the rest frame of the scatterer}. Because of thermal motions, this implies a wavelength shift between the incoming and outgoing photon on top of the Keplerian velocity field. With the thermal motions isotropically and Maxwellian-distributed, the wavelength shift distribution will depend on the angle between the incoming and outgoing direction \citep{Unno1952}. We take this into account by convolving the coherent line luminosity profile with the scattering function, which depends on the angle between the star, the volume element, and the observer. The thermal motions are assumed to correspond to the same broadening parameter $b = 1.5$\,km\,s$^{-1}$ as before.
\end{enumerate}
To illustrate the differences between the three assumptions we plot a projected height emission profile in Fig.~\ref{fig:redist}. Because complete redistribution brings photons in the line wings closer to the line centre (on average), it effectively creates higher optical depths and strongly suppresses the emerging flux from the midplane. For our fiducial model, we used partial redistribution.
\end{enumerate}

\begin{figure}
\centering
   \includegraphics[width=0.5\textwidth]{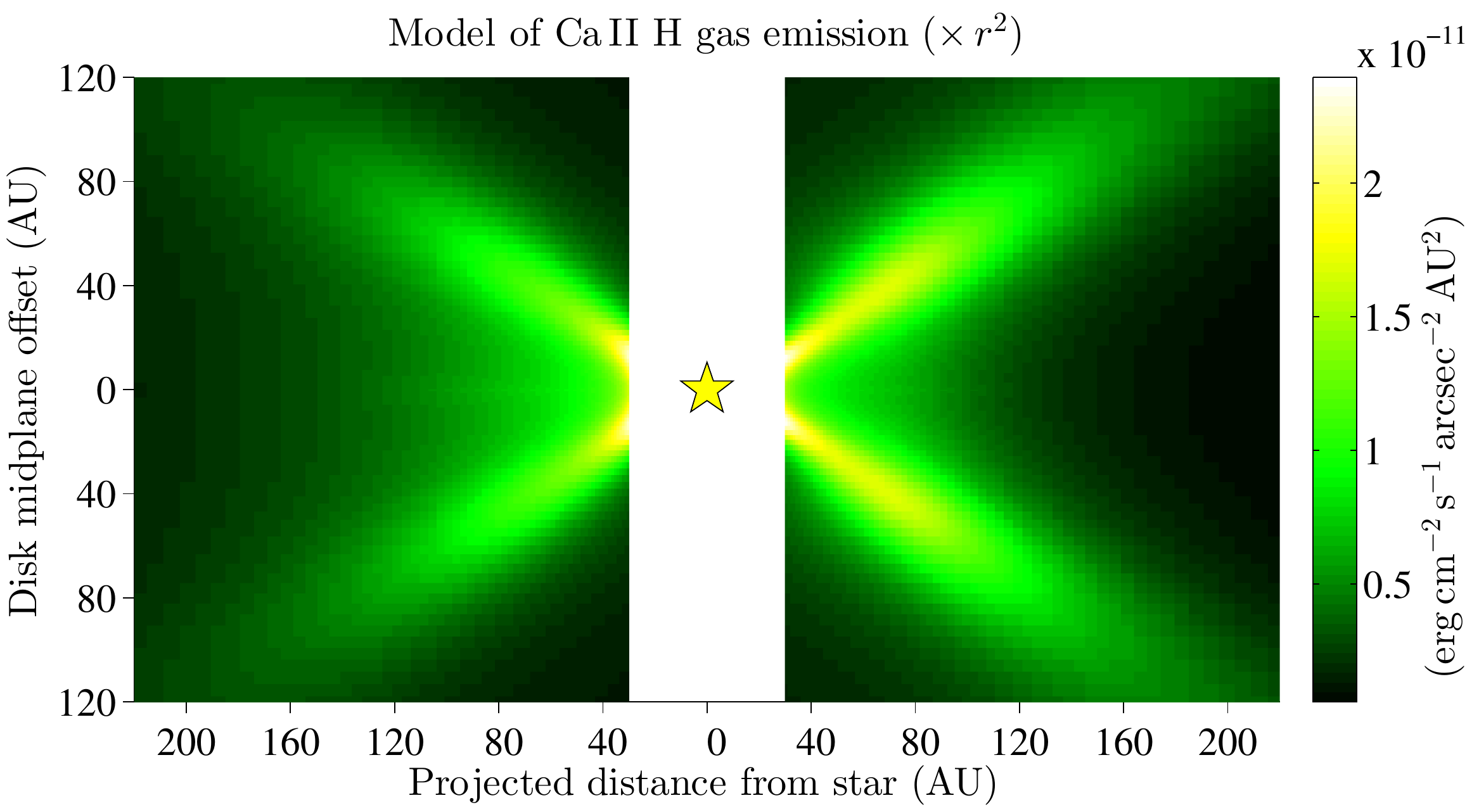}
      \caption{modelled \ion{Ca}{ii} H emission profile, assuming the gas distribution derived from \ion{Fe}{i} $\uplambda$3860.}
      \label{fig:CaII_ontario}
\end{figure}

The resulting \ion{Ca}{ii} H profile from the model (with partial redistribution) is shown in Fig.~\ref{fig:CaII_ontario}, which indeed shows the same characteristics as the observations: suppressed emission from the midplane and emission visible high above the plane. We emphasize that we used precisely the same spatial distribution of gas as was inferred from the optically \ion{Fe}{i} $\uplambda$3860 -- we did not attempt to carefully fit the \ion{Ca}{ii} H emission. Still, the model reproduces the observations closely, but not perfectly. Since the model is symmetric with respect to the midplane, it does not show the strong asymmetries seen in the observations. Furthermore, the emission predicted by the model is about a factor 2--3 higher than seen in the observations. The discrepancies are not too surprising given that we assumed a spatial distribution of gas far beyond where it is well constrained by \ion{Fe}{i} $\uplambda$3860, e.g.\ both much
higher above the midplane and closer in to the star than \ion{Fe}{i} is detected. Small deviations in the \ion{Ca}{ii} density can make big differences in the \ion{Ca}{ii} H line luminosity. This could also explain why the \ion{Ca}{ii} emission extends farther into the SW than the NE, the opposite to what is observed for \ion{Fe}{i}; it could simply be an optical depth effect, through which \ion{Ca}{ii} is shielded by a higher column density towards the star (while the \ion{Fe}{i} is optically thin). Overall, however, we see no evidence that Fe and Ca are \textit{not} well mixed; the striking difference between the observed intensity maps of \ion{Fe}{i} $\uplambda$3860 and \ion{Ca}{ii} H and K is well explained by the difference in optical depths between the lines.

\subsection{Comparison between gas and dust profiles}

\begin{figure}
\centering
   \includegraphics[width=0.5\textwidth]{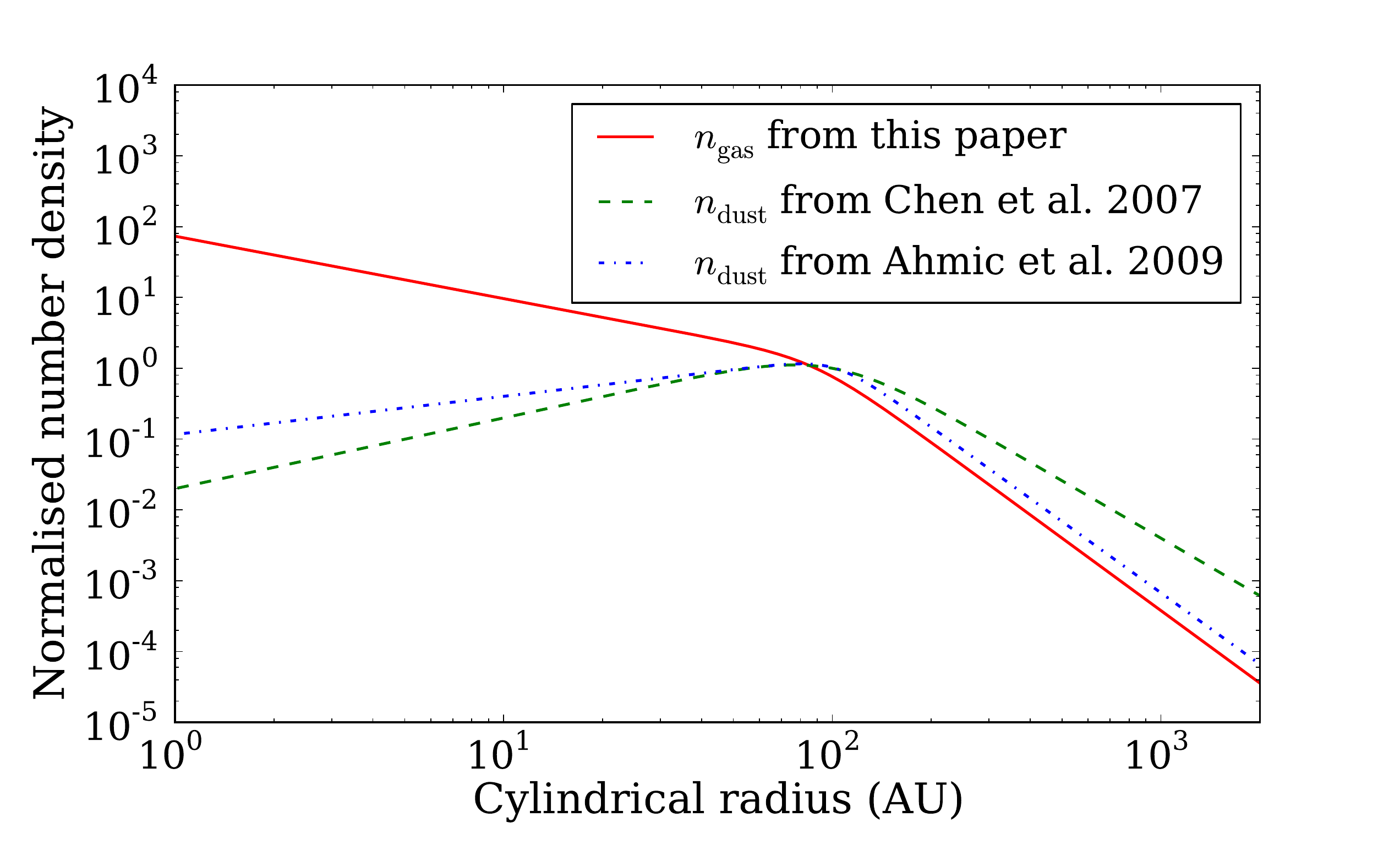}
      \caption{\revone{Number density distribution of gas and dust in the $\beta$\,Pic disk midplane, normalised to unity at the break radius, and shown for mean of parameters derived for NE and SW. We used the one-disk parametric fit from \citet{Ahmic2009}.}}
      \label{fig:gas_vs_dust}
\end{figure}

\revone{The general NE/SW asymmetry of the gas disk is, as shown already by \citet{Brandeker2004}, similar to that of the dust. Our parametric model of the gas number density profile can be compared to the parametric fits of the dust profile by \citet{Chen2007} and \citet{Ahmic2009} to assess how the gas-to-dust ratio varies as a function of disk radius (see Fig.~\ref{fig:gas_vs_dust}). Outside the break radius the distribution of gas follows that of the dust model by \citet{Ahmic2009}, while the results from \citet{Chen2007} indicate a faster decline of the dust number density. Inside 100\,AU the gas number density increases towards the star, in contrast to the dust. Interpreting the connection between the distributions of gas and dust requires more detailed modelling of gas production and redistribution, which is outside the scope of the current paper \citep[see][for a first attempt at relating disk properties to physical production / redistribution mechanisms]{Xie2012}.} 


\section{Summary}\label{sec:con}
Our main results are:
   \begin{enumerate}   
      \item We have produced the first well-sampled \textit{images} of the $\beta$\,Pic gas disk and detected spatially resolved emission from \ion{Fe}{i} and \ion{Ca}{ii}.
      \item We confirm the NE/SW asymmetry previously found by \citet{Brandeker2004}.
      \item We find no evidence for flaring, i.e.\ the disk scale height increases linearly with midplane distance to the star.
      \item The disk height profile drops slower than expected for a gas disk in hydrostatic equilibrium. We speculate that this might be caused by the dynamical interaction between ions and charged dust grains.
      \item \revone{The \ion{Ca}{ii} H- and K spatial emission profiles indicate that Fe and Ca are well mixed throughout the disk; the difference to
      the \ion{Fe}{i} $\uplambda$3860 profile is caused by differences in optical depth.}
   \end{enumerate}


\begin{acknowledgements}
The authors gratefully acknowledge financial contributions from Stockholm Astrobiology Graduate School, and the International Space Science Institute (ISSI) in Bern, Switzerland (ÒExozodiacal Dust Disks and DarwinÓ working group, \texttt{http://www.issibern.ch/teams/exodust/}). A.B.\ was funded by the \emph{Swedish National Space Board} (contract 84/08:1), while K.F.\ was funded by the \emph{Swedish Research Council}. We would also like to thank Reinhard Hanuschik for clarifications regarding the EsoRex pipeline products.
\end{acknowledgements}

\bibliographystyle{aa}
\bibliography{myref} 


\begin{appendix}

\section{Flux calibration and comparison with UVES data}\label{sec:calibcomp}

\begin{figure}[h!]
\centering
   \includegraphics[trim=6mm 3mm 6mm 0mm, clip=true, width=0.5\textwidth]{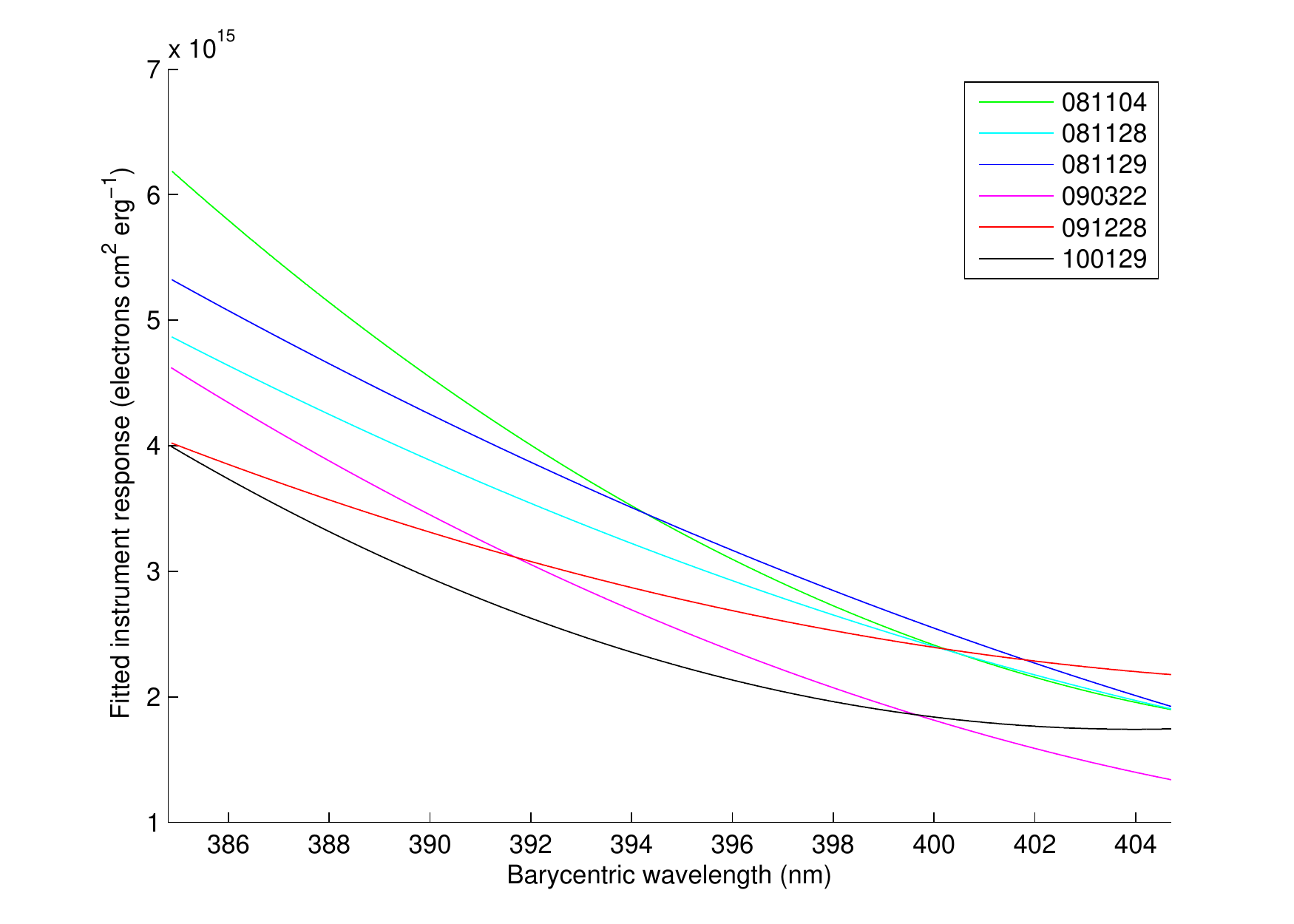}
      \caption{Fitted spectral instrument-response curves from observed standard stars, with colours marking observing dates.}
      \label{fig:instrumentresponse}
\end{figure}

As mentioned in Sect.~\ref{sec:reduction}, the photometric accuracy of FLAMES/GIRAFFE/ARGUS depends to a great extent on the spectral response curve derived from observed standard stars. Unfortunately, the stellar models to which the standards are compared are very coarse in the blue spectral region observed, lacking any detailed spectral features. Moreover, the spectra are extracted by summing up over all pixels in the ARGUS array when running the \emph{EsoRex} pipeline, adding sky emission and cosmic ray spikes. Owing to this, a reasonable spectral instrument-response curve has to be fitted to some well-chosen points in the noisy pipeline-derived instrument response before it can be applied to correct our observed target spectra. The fitted response curves from individual observing nights are presented in Fig.~\ref{fig:instrumentresponse}. Such seemingly large differences in efficiency levels between different observations are unlikely, and must, at least partly, be attributed to photometric uncertainties in the GIRAFFE spectrograph. \revone{Observations of the same standard star at three different occasions (green, blue, and cyan lines), e.g., suggest a spectral instrument-response that varies by as much as 30\% at the bluest wavelengths.} Taking all these factors into account, we estimate a photometric accuracy of at best 20\%.

To examine the relative accuracy of the flux calibration in more detail, we plotted the \ion{Fe}{i} $\uplambda$3860 and the \ion{Ca}{ii} H- and K signal in overlapping inner and outer observed FLAMES regions, as well as overlapping data obtained with UVES by \citet{Brandeker2004}. The \ion{Fe}{i} emission seems to be roughly consistent at higher latitudes in the inner/outer comparison, but there is a clear difference close to the disk midplane, where the strong signal observed in the inner observations appears to be weaker in the outer observations. A similar \ion{Fe}{i} flux deficit of 30--40\% in the outer observations is evident in a comparison with UVES data, but the situation seems to be reversed in the \ion{Ca}{ii} H- and K emission, where the FLAMES fluxes at 60\,AU are $\sim$1.5 times higher than corresponding UVES fluxes, while the data at 120\,AU match fairly well (considering estimated errors for the \ion{Ca}{ii} extraction in that region). A cross-check of the flux-calibrated science spectra with the available HARPS spectrum of $\beta\,$Pic allowed us to confirm that the slope of the fitted response curves is correct to within a few percent, therefore any relative differences between the strength of the \ion{Fe}{i} and \ion{Ca}{ii} emission compared to UVES data is not caused by errors in the fit. Although the error bars assigned to UVES data indicate well-determined fluxes, similar difficulties in determining absolute photometric fluxes exist for that instrument, therefore any calibration of our data with respect to UVES data was not performed. Instead, we assumed that photometric errors and a higher seeing during the outer observations are likely to blame for the observed flux disagreements, and do not affect the qualitative results derived from our modelling (in which relative inner/outer discrepancies are also effectively averaged out).

\begin{figure}
\centering
   \includegraphics[trim=6mm 0mm 6mm 0mm, clip=true, width=0.5\textwidth]{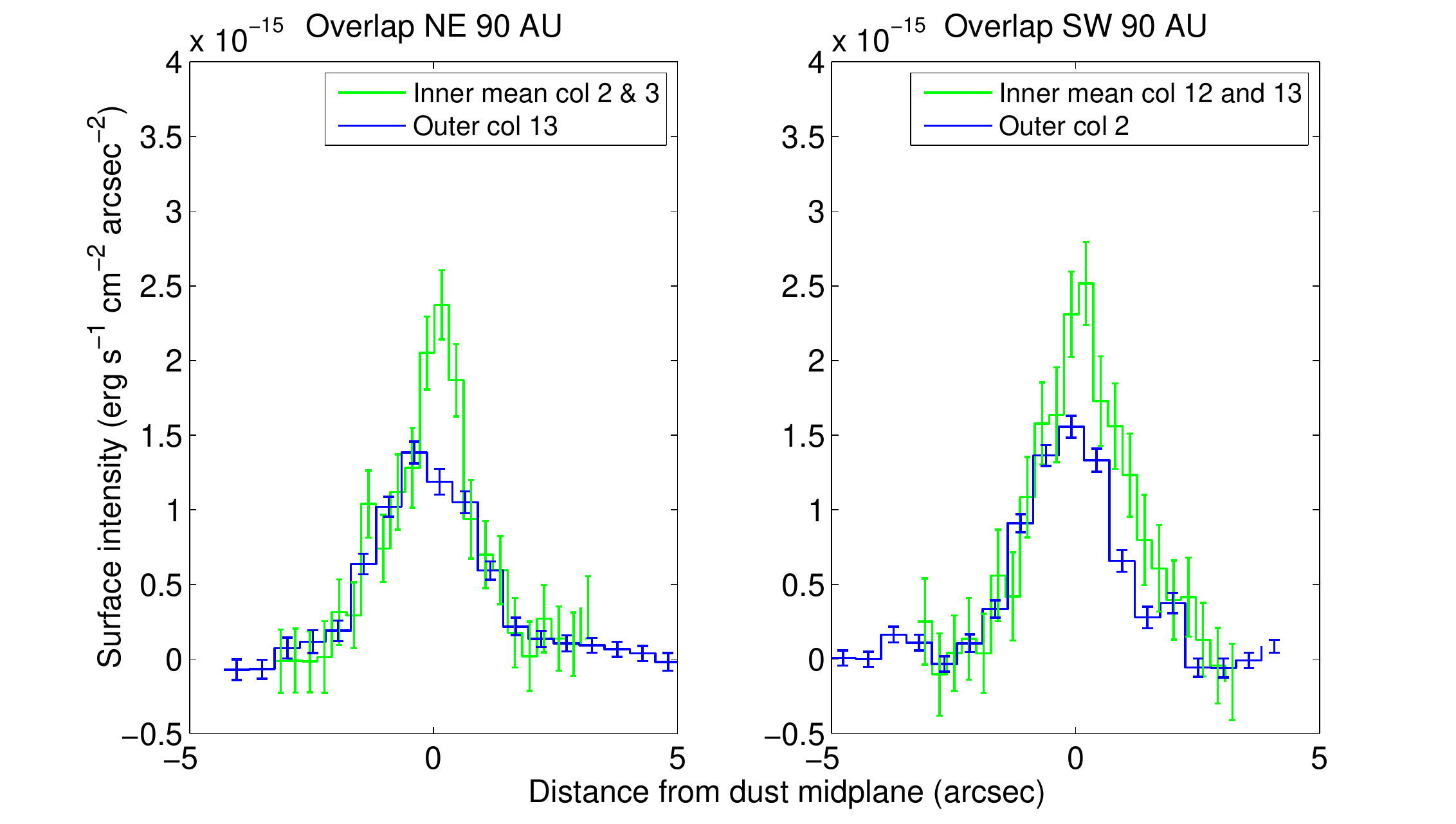}
      \caption{Comparison of \ion{Fe}{i} emission in overlapping areas of inner and outer NE and SW observations. The pixel columns are counted from left to right in the four observed regions.}
      \label{fig:overlapcomp_FeI}
\end{figure}

\end{appendix}

\listofobjects

\end{document}